\documentclass[fleqn,usenatbib]{mnras}

\usepackage{newtxtext,newtxmath}

\usepackage{ae,aecompl}

\usepackage{graphicx}	
\usepackage{amsmath}	
\usepackage{amssymb}	
\usepackage{multicol}
\usepackage{hyperref}
\usepackage{enumitem}

\title[Lockman Hole 325 MHz]{Characterizing EoR foregrounds: A study of the Lockman Hole Region at 325 MHz}

\author[A Mazumder et al.]{
Aishrila Mazumder$^{1}$\thanks{E-mail: aishri0208@gmail.com}
Arnab Chakraborty,$^{1}$
Abhirup Datta,$^{1}$
Samir Choudhuri,$^{2}$ 
\newauthor
Nirupam Roy,$^{3}$
Yogesh Wadadekar,$^{4}$
C. H. Ishwara-Chandra$^{4}$
\\
$^{1}$Discipline of Astronomy, Astrophysics and Space Engineering, Indian Institute of Technology Indore, Indore 453552, India\\
$^{2}$School of Physics and Astronomy, Queen Mary University of London, London E1 4NS, UK\\
$^{3}$Department of Physics , Indian Institute of Science,Bangalore 560012,India\\
$^{4}$ National Centre for Radio Astrophysics, Tata Institute of Fundamental Research, Post Bag 3, Ganeshkhind, Pune 411007, India
}
\date{Accepted XXX. Received YYY; in original form ZZZ}

\pubyear{2015}

\begin{document}
\label{firstpage}
\pagerange{\pageref{firstpage}--\pageref{lastpage}}
\maketitle

\begin{abstract}
One of the key science goals for the most sensitive telescopes, both current and upcoming, is the detection of the redshifted 21-cm signal from the Cosmic Dawn and Epoch of Reionization. The success of detection relies on accurate foreground modeling for their removal from data sets. This paper presents the characterization of astrophysical sources in the Lockman Hole region. Using 325 MHz data obtained from the GMRT, a $6^\circ \times 6^\circ$ mosaiced map is produced with an RMS reaching 50 $\mu$Jy $\mathrm{beam}^{-1}$. A source catalog containing 6186 sources is created, and the Euclidean normalized differential source counts have been derived from it, consistent with previous observations as well as simulations. A detailed comparison of the source catalog is also made with previous findings - at both lower and higher frequencies. The angular power spectrum (APS) of the diffuse Galactic synchrotron emission is determined for three different galactic latitudes using the Tapered Gridded Estimator. The values of the APS lie between $\sim$1 mK$^2$ to $\sim$100 mK$^2$. Fitting a power law of the form $A\ell^{-\beta}$ gives values of  $A$ and $\beta$ varying across the latitudes considered. This paper demonstrates, for the first time, the variation of the power-law index for diffuse emission at very high galactic locations. It follows the same trend that is seen at locations near the galactic plane, thus emphasizing the need for low-frequency observations for developing better models of the diffuse emission.  
\end{abstract}




\section{Introduction}

The Cosmic Microwave Background (CMB), the afterglow from the Big Bang, has provided unprecedented insights into the thermal history of the Universe dating back to the surface of last scattering. However, despite the development of precision instruments for cosmology, the period after recombination (Dark Ages) of the Universe, till the neutral Universe became ionized once again remains one of the relatively unexplored realms in cosmology. It is during this time, between redshifts $\sim$ 30-12, that tiny fluctuations in matter density grew under gravitational instabilities that resulted in the generation of the first stars. This epoch is known as the Cosmic Dawn (CD) - the dawn for the luminous sources in the Universe. This first generation of stars produced a copious amount of ionizing radiation (X-rays, UV and Ly$\alpha$), which converted the neutral intergalactic medium into an ionized one \citep{loeb2001}. This era is known as the Epoch of Reionization (EoR) (see \citealt{Furlanetto2006, morales2010, Pritchard2012} for comprehensive reviews). Observations of quasar absorption spectra \citep{fan2006}, as well as Thompson optical depth obtained from CMB temperature and polarization measurements \citep{Planck2018I},constrain the EoR to $6 < z < 15$. 

To investigate the physical processes occurring in these early stages of the Universe, the most promising probe is the 21-cm hyperfine line of neutral hydrogen \citep{Field1958, Field1959a, Field1959b}. This signal from the early Universe is measured from its temperature contrast against the CMB temperature \citep{Madau1997}. Various physical processes occurring in the early Universe determine whether the signal is detected in absorption or emission. Since the signal evolves with redshift (or frequency), detecting its variation with redshift is equivalent to capturing various properties of the Intergalactic Medium (IGM) during these epochs. For observational detection of the signal, mainly two approaches are taken - observation of all-sky averaged 'global' signal and observing its tiny statistical fluctuations. Several ongoing, as well as next-generation radio astronomical interferometers, are targeting the power spectrum measurements as well as tomographic imaging of the IGM during these early epochs. Instruments that are currently operational like GMRT, LOFAR, MWA have so far placed upper limits on the brightness temperature distribution of the 21-cm signal from reionization \citep{pacgia, mwa1, LOFAR}. There are several dedicated instruments for the detection of the 21-cm fluctuations from CD and EoR. This includes the Donald C. Backer Precision Array to Probe the Epoch of reionization (PAPER, \citealt{Parsons_2010}), the Hydrogen Epoch of Reionization Array (HERA, \citealt{DeBoer_2017}); also, the MWA Phase 2 upgrade \citep{wayth_2018} includes 72 new short spaced tiles for EoR science. Recent upper limits to the EoR power spectrum have been placed from PAPER \citep{Kolopanis_2019} and MWA \citep{li2019season}. However, several factors have so far prevented the actual detection of the power spectrum of the 21-cm emission from CD/EoR.  

 The key factors responsible for obscuring the signal are instrument chromaticity, the precision of data calibration, and the presence of bright foregrounds (compact as well as diffuse). The foreground sources include diffuse emission like synchrotron emission from the galaxy \citep{Shaver1999} as well as from low redshift clusters \citep{dimatteo2004},  free-free emissions from both the galaxy as well as extragalactic sources \citep{Cooray2004} and emission from faint radio-loud quasars \citep{DiMatteo2002}. Point sources are also a significant contributor to the contaminants \citep{Datta2009}. These foregrounds have amplitudes, which are 4-5 orders of magnitude higher than the redshifted 21-cm signal along any line of sight \citep{Zaldarriaga2004, bharadwaj2005, jelic2008, jelic2010, Chapman2015}. Recovering the faint signal amidst the bright sea of foregrounds requires precision instrumentation as well as sophisticated algorithms. However, any strategy used is based on the one property of the foregrounds - their spectral smoothness. The 21-cm has a spectral shape in contrast to the foregrounds, which are assumed to be spectrally smooth \citep{Pritchard2012}. This property of the foregrounds can be exploited to extract the signal of interest. There are three primary strategies for handling foregrounds - foreground avoidance, foreground suppression, and foreground removal \citep{Datta2010, Trott2012, Chapman2015, Chapman2016}. The redshifted 21-cm signal, as opposed to the spectrally smooth foregrounds, show rapid decorrelation over a frequency separation of $\sim$ 1 MHz. Hence, the isolation of the cosmological signal from the foregrounds (see \citealt{ghosh2011} and references therein) may be plausible. Thus, although the signal of interest is extremely faint, with sensitive enough telescopes and a careful investigation of the foregrounds, it is recoverable. With the inception of the Square Kilometer Array (SKA), the expected observational sensitivity would be sufficient to detect the signal statistically \citep{Mellema2014, Koopmans2015}. However, since there is no full-proof strategy for dealing with foregrounds (all three of the above methods come with both advantages and shortcomings), it is still a trial-and-error quest for the telescopes to determine the best strategy for signal extraction given the bright foregrounds.

Nevertheless, for the determination of a perfect strategy, it is essential to have enough observational data to produce accurate models of the foreground sources. A number of models have predicted that the frequency range of the CD and EoR observation would lie between $\sim$ 50 MHz to $\sim$ 200 MHz, since the redshift range of interest is expected to lie between $6 < z < 30$ \citep{Pritchard2012, Loeb2013}. At these frequencies, the dominant foreground is the Diffuse Galactic Synchrotron Emission (DGSE). It dominates at larger angular scales (or smaller baselines) where the EoR signal is also expected to have the maximum sensitivity. The longer baselines (or smaller angular scales) have major foreground contribution from extragalactic compact source populations. The discrete source population is mainly dominated by various classes of Active Galactic Nuclei (AGNs) and Star-Forming Galaxies (SFGs). For accurate foreground modeling, the nature and population of these sources in terms of their spatial, as well as spectral characteristics, become essential. Hence, low-frequency all-sky observation to determine the discrete source distribution becomes necessary. Using available data from deep field observations as well as physical models, various state of the art simulations are being performed to simulate source population and flux distribution at low frequency for characterizing foregrounds \citep{DeOliveira-Costa2008, S3, trecs}. However, for more accurate predictions, more observational evidence is required. In addition to spectral smoothness of the foregrounds, their angular power spectrum can also be exploited for characterizing the sky signal, which consists of the 21-cm signal together with foregrounds \citep{Datta2010,ghosh2010,sims2016}. The power spectrum of DGSE is modeled in a power-law form as a function of frequency and angular scale \citep{Santos2005, kanan_2007, ali08}, expressed as :

\begin{equation}
\mathcal{C}_{\mathcal{\ell}}(\nu) = A \Big(\frac{\mathcal{\ell}}{\mathcal{\ell}_{0}}\Big) ^{ -\beta} \Big(\frac{\nu}{\nu_{0}}\Big)^{-2\alpha}, 
\label{power_law_model}
\end{equation}
with $\beta$ as the power-law index of the angular power spectrum of DGSE and $\alpha$ as the mean spectral index. For the foreground subtraction method, a power-law fit for the DGSE is done for each pixel along the frequency direction in the data cube and hence subtracted from the data \citep{jelic2008,liu09, Datta2009,simspober2019}. But precise modeling of this power law is required to prevent removal of the signal along with the foreground. 

Low-frequency observation of radio sky is also essential for the study of the astrophysics at play in various evolutionary stages of different galactic and extragalactic sources. Radio emission at low frequencies, together with their redshift information, can be used to infer several astrophysical properties associated with the sources. In general, the source distribution is assumed Poissonian (with the possibility of clustering following single power-law) \citep{ali08, jelic2010, Trott_2016}. Source counts are also modeled via a single power-law distribution \citep{Intema16, gleam, franzen2019}. However, several recent studies have shown deviation from the single power-law model \citep{williams2016,prandoni2018, Hale19}. Thus more detailed studies both in wide-field as well as deep fields are required for generating a fiducial model of sources at low frequencies. The differential source counts at these are also useful for constraining the nature of sources. At frequencies in the GHz range and upwards, the source properties are well characterized. However, at lower frequencies, there is a lack of consensus for the same. It has been seen in previous studies that at these frequencies, AGNs dominate in the flux density scales down to few 100 $\mu$Jy while SFGs and radio quite AGNs become dominant below $\sim$ 100 $\mu$Jy \citep{Simpson06, seymour08, mignano, Smolic08, Padovani2009, Padovani2011, padovani2015,prandoni2018, Hale19}. This is inferred from the flattening of the source counts below 1 mJy. However, such studies are very few on account of the limitations in reaching the required SNR. Thus empirical constraints at frequencies $\lesssim ~1.4$ GHz are limited. Therefore the study of the low-frequency radio sky is vital for fiducial modeling of foregrounds for 21-cm cosmology as well as constraining the physics and the astrophysics of the sources.    

 This work presents the low-frequency properties of an extragalactic region at high galactic latitude. The field studied is the Lockman Hole region \citep{Lockman1986}, the area with the lowest HI column density in the sky, having a low infra-red background \citep{Lonsdale_2003}. Lockman Hole is one of the most extensively studied extragalactic fields, with a large number of studies in optical, X-ray, UV, and IR bands. Many studies of the region also exists in the radio frequencies, but mostly upwards of 1.4 GHz \citep{prandoni2018,vernstrom2016a,vernstrom2016b,vernstrom14,Condon_2012,ibar09, Owen_2009,Owen_2008,ciliegi, deRuiter1997}. At lower frequencies, observations have been done at 610 MHz by \citet{Garn08,Garn2010} and at 150 MHz by \citet{mahony2016}. There is a requirement for more observational analyses of this region at the lower frequency end to complement the pre-existing analyses.
 The Lockman Hole field, by virtue of its high galactic latitude, is very well suited for constraining the DGSE power spectrum (since the EoR target fields selected for current and upcoming telescopes are at high galactic latitudes).

 In this work, 325 MHz data of the Lockman Hole region from the legacy GMRT is analyzed. At this observing frequency, the only other observation in literature is the VLA observation of \citet{Owen_2009}. For that work, 324.5 MHz observation with a $\sim$ 2.3$^\circ$ FoV centered at $10^{h}06^{m}01^{s}$  $+34^{d}54^{m}10^{s}$ was considered. The RMS sensitivity reached was 70 $\mu$Jy $\mathrm{beam}^{-1}$. For this work, a 6$^\circ$ $\times$ 6$^\circ$ map of the field reaching an RMS level of 50 $\mu$Jy $\mathrm{beam}^{-1}$ has been produced. Due to the large field of view covered (because of multiple pointings), the data has significant coverage, which enables the study of the variation of the DGSE power spectrum with latitude. A source catalog, containing sources with fluxes down to 250 $\mu$Jy containing 6186 sources has also been generated. 

 The catalog thus produced has been compared with previous catalogs (mostly at higher frequencies) for the same region. Euclidean normalized source count has also been determined and compared to prior studies. The Tapered Gridded Estimator (TGE, \citealt{tge14, tge16}) has been used to determine the angular power spectra for the DGSE using three separate pointings (within the observed field of view) located at three different galactic latitudes. This has been done to study the variation in the nature DGSE with increasing distance from the galactic center. The Lockman Hole region, despite having an extensive multi-wavelength coverage, nevertheless has some gap in the lower frequency end. However, for the characterization of extragalactic sources and determination of the astrophysics at play,  more multi-wavelength coverage is essential. This paper aims to fill the low-frequency observational gap for better multi-frequency source characterization of astrophysical sources. 

The paper starts by describing the observation and the data reduction process in section \ref{sec.observation}. Section \ref{sec.source_catalog} details out the sources extracted and cataloged after reducing the data. In section \ref{sec.comparison}, comparison with existing observations has been discussed. The flux for the sources extracted from the cataloged is corrected for various factors and binned to obtain the Euclidian normalized source counts - which is described in section \ref{sec.dnds}. Section \ref{sec.DGSE} describes the variation of the angular power spectrum across the field. Finally, section \ref{conclusion} concludes the paper with a discussion of the implications of this work and the further studies required.

\section{Observation and Data Reduction}
\label{sec.observation}
 
This work uses the archival data of the Lockman Hole observed by Giant Metrewave Radio Telescope (GMRT, \citealt{Swarup1991}). 
The observations were carried out during cycle 23 (Project Code - 23$\_$001) with the legacy GMRT (GMRT Software Backend; GSB) at 325 MHz frequency band with an instantaneous bandwidth of 32 MHz. Details of the observing parameters are outlined in Table \ref{observation}. Data were taken over 11 nights between February 15 and March 20, 2013, and has 23 different pointing centers. The field center of entire mosaic of 23 pointings is at ($\alpha_{2000}=10^{h}48^{m}00^{s},\delta_{2000}=58^{\circ}08'00\arcsec$ ). Two flux calibrators- 3C48 and 3C286 are observed respectively at the beginning and end of each night's observation because the former one would have set at the time each of the observing run ended. Between each set of target observations (each set is the successive observation of the 23 targeted pointings), there is an intermittent observation of the source 1006+349, which is the phase calibrator. 

The choice of the above data is due to several interesting features. It is one of the very few data having a large field of view coverage at such low frequencies. There is also a lack of study at the frequency considered here. This region, being located quite far from the galactic plane (phase center is at $l=149.18^\circ, b=52.24^\circ$), presents an intriguing field for study and characterization of extragalactic astrophysical sources.

\begin{table}
\caption{Observational details of the target field and the calibrator sources for all the observing sessions}
	
\begin{tabular}[width=\columnwidth]{ll}
\hline
\hline
Project code & 23\_001$^\ddagger$\\
Observation date & 15 \& 18 February 2013 \\
                 &  4,5,6,7,11,12,13,14,20 March 2013\\
(as obtained from data) & \\                 
\hline
Bandwidth &  32 MHz\\
Frequency range & 309-341 MHz\\
Channels & 256\\
Correlations & RR RL LR LL\\
Flux calibrator & 3C48 \& 3C286 \\
Phase calibrator & 1006+349 \\
\hline
Pointing centres & $01^{h}37^{m}41^{s}$ $+33^{d}09^{m}35^{s}$ (3C48)\\
                 & $10^{h}06^{m}01^{s}$  $+34^{d}54^{m}10^{s}$ (1006+349) \\
                 & $11^{h}03^{m}18^{s}$  $+59^{d}44^{m}50^{s}$ (Target) \\
                 & $10^{h}55^{m}54^{s}$  $+59^{d}44^{m}50^{s}$ (Target) \\
                 & $10^{h}48^{m}29^{s}$  $+59^{d}44^{m}50^{s}$ (Target) \\
                 & $10^{h}41^{m}05^{s}$  $+59^{d}44^{m}50^{s}$ (Target) \\
                 & $10^{h}33^{m}41^{s}$  $+59^{d}44^{m}50^{s}$ (Target) \\
                 & $10^{h}59^{m}36^{s}$  $+58^{d}53^{m}55^{s}$ (Target) \\
                 & $10^{h}52^{m}12^{s}$  $+58^{d}53^{m}55^{s}$ (Target) \\
                 & $10^{h}44^{m}47^{s}$  $+58^{d}53^{m}55^{s}$ (Target) \\
                 & $10^{h}37^{m}23^{s}$  $+58^{d}53^{m}55^{s}$ (Target) \\
                 & $11^{h}03^{m}18^{s}$  $+58^{d}03^{m}00^{s}$ (Target) \\
                 & $10^{h}55^{m}54^{s}$  $+58^{d}03^{m}00^{s}$ (Target) \\
                 & $10^{h}48^{m}29^{s}$  $+58^{d}03^{m}00^{s}$ (Target) \\
                 & $10^{h}41^{m}05^{s}$  $+58^{d}03^{m}00^{s}$ (Target) \\ 
                 & $10^{h}33^{m}41^{s}$  $+58^{d}03^{m}00^{s}$ (Target) \\
                 & $10^{h}59^{m}18^{s}$  $+57^{d}12^{m}04^{s}$ (Target) \\
                 & $10^{h}52^{m}12^{s}$  $+57^{d}12^{m}04^{s}$ (Target) \\
                 & $10^{h}44^{m}47^{s}$  $+57^{d}12^{m}04^{s}$ (Target) \\
                 & $10^{h}37^{m}23^{s}$  $+57^{d}12^{m}04^{s}$ (Target) \\
                 & $11^{h}03^{m}18^{s}$  $+56^{d}21^{m}09^{s}$ (Target) \\ 
                 & $10^{h}55^{m}54^{s}$  $+56^{d}21^{m}09^{s}$ (Target) \\ 
                 & $10^{h}48^{m}29^{s}$  $+56^{d}21^{m}09^{s}$ (Target) \\ 
                 & $10^{h}41^{m}05^{s}$  $+56^{d}21^{m}09^{s}$ (Target) \\ 
                 & $10^{h}33^{m}41^{s}$  $+56^{d}21^{m}09^{s}$ (Target) \\ 
                 & $13^{h}31^{m}08^{s}$  $+30^{d}30^{m}32^{s}$ (3C286)\\
                 
\hline                 
\hline                 
\end{tabular}
\label{observation}	
\flushleft{${\ddagger}$ as given in the cover-sheet accompanying the data}\\
\end{table}

\subsection{Data Reduction}

Data were reduced using the Source Peeling and Atmospheric Modelling (SPAM) pipeline \citep{Intema2009, Intema2014s, Intema2014, Intema16}. It is a fully automated software based on the Astronomical Image Processing System (AIPS, \citealt{Greisen1998, Greisen2002}).
The primary rationale of using SPAM is that it does a "direction-dependent" calibration, besides the traditional "direction-independent" calibration. Direction dependent approach of SPAM iteratively solves for ionospheric phases and hence corrects for ionospheric phase errors. Such phase errors are expected to be persistent, given the large field of view as well as low frequencies used for observing the current data. Therefore, modeling and correction of the ionospheric dispersive delays improve the background noise and flux-scale accuracy. The data reduction steps have been described in brief below:

\begin{enumerate}[label=(\alph*)]

    \item \textbf{Pre-processing}: Initially, SPAM computes the instrumental calibration using the best available scan of the primary calibrator and calibrates the data using these solutions. The flux densities of the primary calibrators are set following the \citet{scafie-heald2012} flux density scale. After the initial RFI flagging and bad data removal, both the time-varying complex gain solutions and time constant bandpass solutions are computed per antenna per polarization. In the pre-processing step, data is also averaged in time and frequency to reduce data volume for subsequent calibration steps.
    
    \item \textbf{Direction Independent Calibration}: This step is akin to the classical self-calibration. It uses the NVSS derived sky model for performing the self-calibration on the data. The calibrated visibilities are imaged with wide-field imaging using Briggs weighting with robust parameter -1 (which produces very well-behaved point spread functions by down-weighting the dense central UV plane coverage of GMRT). 
    
    \item \textbf{Direction Dependent Calibration}: Gain phases and the sky model generated in the previous step are used to initiate the direction-dependent calibration. SPAM determines the gain phases by peeling the apparently bright sources in the field. The phase correction factors generated by this process is a measure of the ionospheric phase delay. These solutions are applied to the data calibrated in the previous step. It creates of visibilities with correction for ionospheric phases. At the end of the direction-dependent step, the primary beam corrected image is obtained. 
    

\end{enumerate}

\begin{figure*}

\includegraphics[width=7.0in,height=6.0in]{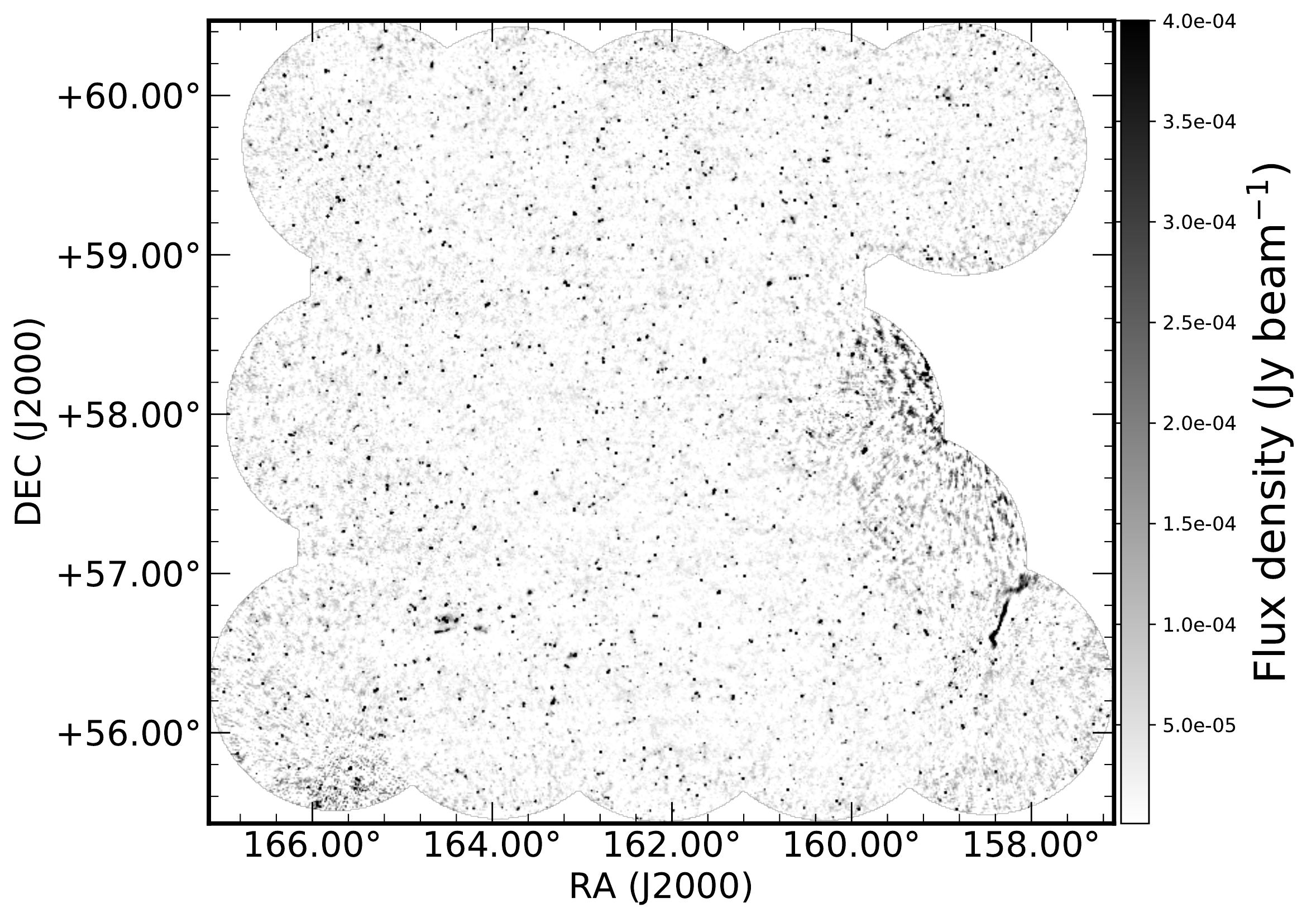}

 \caption{Primary beam corrected mosaic of the Lockman Hole region at 325MHz. The off source RMS at the center is $\sim$ 50 $\mu$Jy $\mathrm{beam}^{-1}$ and beam size is $ 9.0\arcsec \times 9.0\arcsec$}
\label{PB}
\end{figure*}

\begin{figure*}
\includegraphics[width=4.5in,height=3.5in]{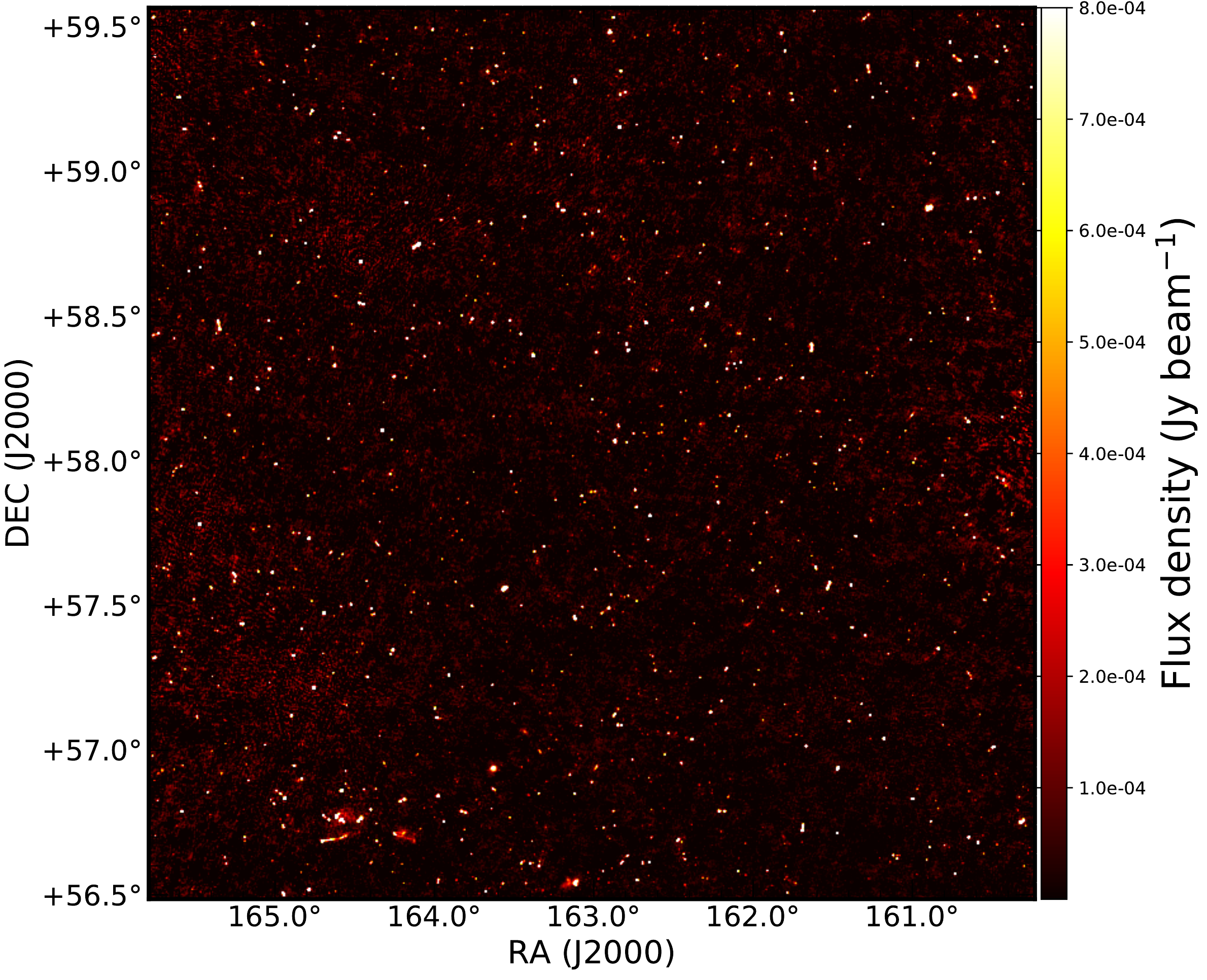}
\caption{Zoomed-in Stokes I image of the central region of Lockman Hole at 325MHz, covering an area of $\sim$ 5 $\mathrm{deg}^{2}$. The Central off-source noise is $\sim$ 50 $\mu \mathrm{Jy}$ $\mathrm{beam}^{-1}$}
\label{image}
\end{figure*}

For each observing night, the archival data is downloaded in LTA format, together with a FLAGS file containing all the information regarding the telescope system during the observation. SPAM was used to pre-calibrate the data for each night's observation. 
The individual pointings were processed separately, post concatenation of the data for all nights. Running the pipeline, as mentioned above, gives calibrated visibilities, which are both time and channel averaged. The visibility files thus produced consist of calibrated uvfits files (containing calibrated UV data) and residual uvfits files (containing residual visibilities). 
The averaged data has 42 channels, with width 781.25 KHz, i.e., the effective bandwidth remains 32 MHz. Each of such pointings was individually imaged, to produce primary beam corrected image. The primary beam corrected images of each pointing were then mosaiced to create the final image. The image is $6^\circ \times 6^\circ$, as shown in Figure \ref{PB}. It is to be mentioned that two pointings (located off the center) had to be removed due to the presence of substantial noise in the final PB-corrected image. In Figure \ref{image}, the zoomed-in image of the phase center is shown. The off-source noise is $\sim50\mu$Jy $\mathrm{beam}^{-1}$, covering almost $6^\circ \times 6^\circ$   with a beam size of $9\arcsec \times 9\arcsec$.   

 \section{Source Catalog}
 \label{sec.source_catalog}

 A source catalog was produced using Python Blob Detection and Source Finder \footnote{\url{https://www.astron.nl/citt/pybdsf/}}(P{\tiny Y}BDSF, \citealt{Mohan2015}) for characterization of the sources present in the field. P{\tiny Y}BDSF also produced a residual map, which is the image of the field after subtracting all the modeled point sources. The source catalog was generated using the primary beam corrected mosaiced image of the field. P{\tiny Y}BDSF uses a sliding box window to calculate the RMS variation across the field. For this work, a box of size 120 pixels every 30 pixels was used, i.e., rms$\_$box = (120,30). Using a threshold of 100$\sigma_{\mathrm{RMS}}$, bright regions with peak amplitude higher than this value were first isolated (where$\sigma_{\mathrm{RMS}}$ is the clipped RMS across the entire map). A smaller box of size rms$\_$box$\_$bright = (30,10) was used around the bright regions to avoid considering artifacts around these regions as real sources. It identifies islands that have contiguous emission and then fits multiple Gaussians to each island. The selection threshold for islands is $3\sigma_{\mathrm{RMS}}$ and the same for source detection is $6\sigma_{\mathrm{RMS}}$. 
 
 The sources are identified by P{\tiny Y}BDSF by grouping neighboring Gaussians into sources. The total flux of a source is the sum of all fluxes in a group of Gaussians. The flux uncertainty is the quadrature sum of individual uncertainties of the Gaussians. The source position is the centroid, while the spatial size is determined using moment analysis using restoring beam size. 
 
\begin{figure*}
 \centering
\includegraphics[width=\columnwidth,height=3.0in]{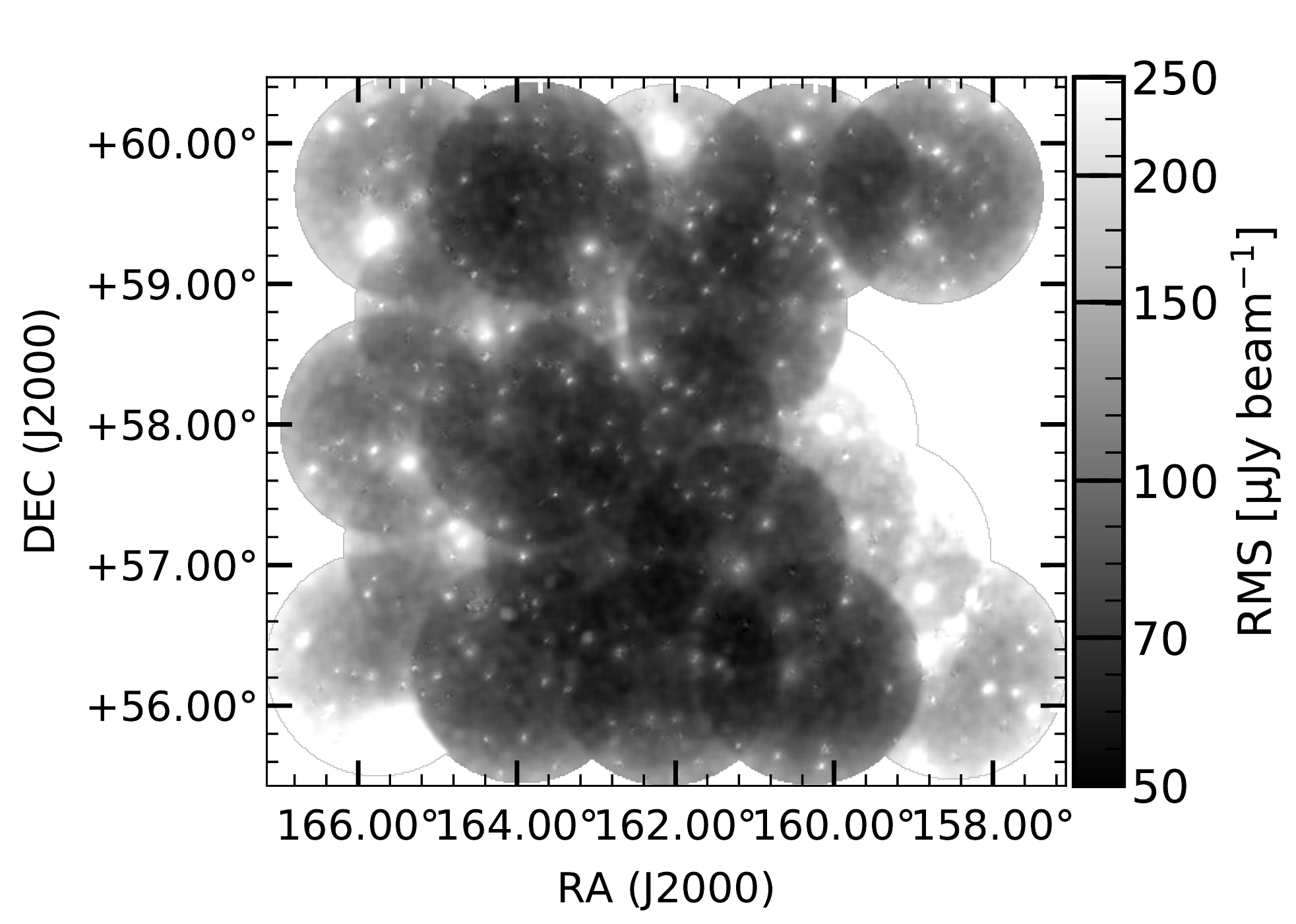} 
 \includegraphics[width=\columnwidth,height=2.9in]{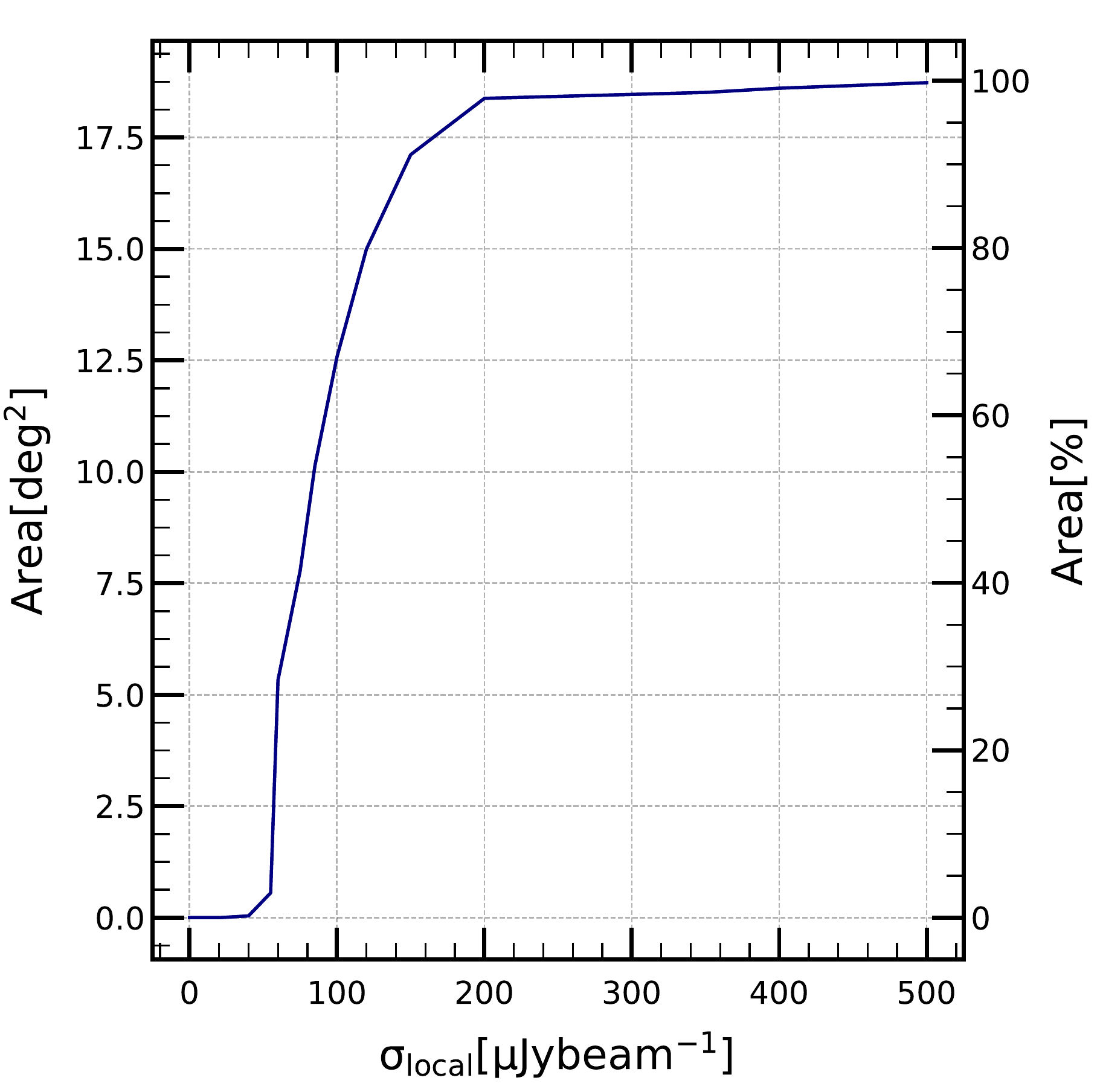}
\caption{(\textbf{Left}) Local RMS noise in the final mosaic; (\textbf{Right}) Cumulative area of the final mosaic as a function of the RMS noise}
\label{rms}
\end{figure*}

 The ionospheric fluctuation over the large field of view, given the low observing frequency, would considerably vary the PSF. Thus the actual PSF may be slightly different than the restoring beam at different parts of the image. Setting the option $psf\_vary\_do = \mathrm{True}$\ in P{\tiny Y}BDSF takes care of the PSF variation across the field of view (readers are referred to the P{\tiny Y}BDSF documentation available from the link in footnote 1 for more details). 
 An RMS map was also produced that depicted variation of the background noise across the field. As can be seen in Figure \ref{rms} (left), the background RMS is quite high near the bright sources and particularly high towards the edges of the field. 
 
Every source catalog generated, besides being resolution and thermal noise limited, is also limited by confusion noise. Confusion noise is defined as the background sky brightness fluctuation due to several faint sources in one telescope beam.

Using the formula described in \citet{Condon_2012}, 
 \begin{equation}
     \sigma_c = 1.2 \mathrm{\mu Jy beam^{-1}}(\frac{\nu}{3.02 GHz})^{-0.7}(\frac{\theta}{8^{\arcsec}})^{10/3}
 \end{equation}
(where $\nu$ is observing frequency, and $\theta$ is the FWHM of the telescope beam\footnote{\url{https://www.cv.nrao.edu/course/astr534/Radiometers.html}}) the confusion noise limit is 8.46 $\mu$Jy $\mathrm{beam}^{-1}$.

The catalog assembled in this work comprises of 6186 sources above the $5\sigma_{RMS}$ threshold. A sample of the catalog is shown in Table \ref{sample_catalog} (full catalogue is available with the electronic version of the paper).

\begin{table*}
\caption{Sample of 325 MHz source catalog of  Lockman Hole region generated}
\label{sample_catalog}
\scalebox{1.0}{
\begin{tabular}{l l l l l l l l l l l l}
\hline
\hline
source{\textunderscore}id & RA & E{\textunderscore}RA & DEC & E{\textunderscore}DEC & Total{\textunderscore}flux & Peak{\textunderscore}flux & RMS & Maj &  Min & PA \\
() & (deg) & (arcsec) & (deg) & (arcsec) & (mJy) & (mJy $\mathrm{beam}^{-1}$) & (mJy $\mathrm{beam}^{-1}$) & (arcmin) & (arcmin) & (deg) \\
 \hline
0. & 167.2262 & 0.0957 & 56.5286 & 0.1377 & 11.99 & 9.11 & 0.23 &  0.18 & 0.14 & 115.13 \\
1. & 167.1580 & 0.3182 & 56.2800 & 0.2681 & 4.39 & 3.55 & 0.23 & 0.19 & 0.15 & 63.28 \\
2. & 167.1452 &  0.3847 & 56.3989 & 0.4378 & 3.85 & 2.67 & 0.22 & 0.2 & 0.17 & 36.15 \\
3 .& 167.0852 & 0.0809 & 56.6260 & 0.0606 & 42.18 & 23.27 & 0.26 & 0.23 & 0.18 & 25.85 \\
4. & 167.0834 & 0.2681 & 56.2796 & 0.2483 & 5.37 & 3.96 & 0.21 & 0.19 & 0.16 & 133.3 \\
5. & 167.3339 & 1.1565 & 58.0057 & 0.6568 & 1.81 & 1.09 & 0.18 & 0.23 & 0.16 & 108.93 \\
6. & 167.3102 & 0.1834 & 58.6486 & 0.1009 & 12.32 & 6.94 & 0.17 & 0.24 & 0.17 & 87.42 \\
7. & 167.0687 & 0.0923 & 56.6531 & 0.0887 & 14.09 & 11.2 & 0.23 & 0.17&  0.16 & 61.47 \\
8. & 167.0688 & 0.1891 & 58.1628 & 0.2458 & 12.15 & 6.15 & 0.23 & 0.23 & 0.19 & 24.84 \\ 
9. & 167.2761 & 0.1507 & 56.0524 & 0.2582 & 9.51 & 4.64 & 0.16 & 0.25 & 0.18 & 2.54 \\
\hline
\end{tabular}}
\\\flushleft{Notes: The columns of the final catalog (fits format) include source ids, positions, error in positions, flux densities, peak flux densities, local RMS noise, sizes, and position angle respectively.}
\end{table*}
 
 \subsection{Classification}

\begin{figure}
\centering
\includegraphics[width=\columnwidth, height=3in]{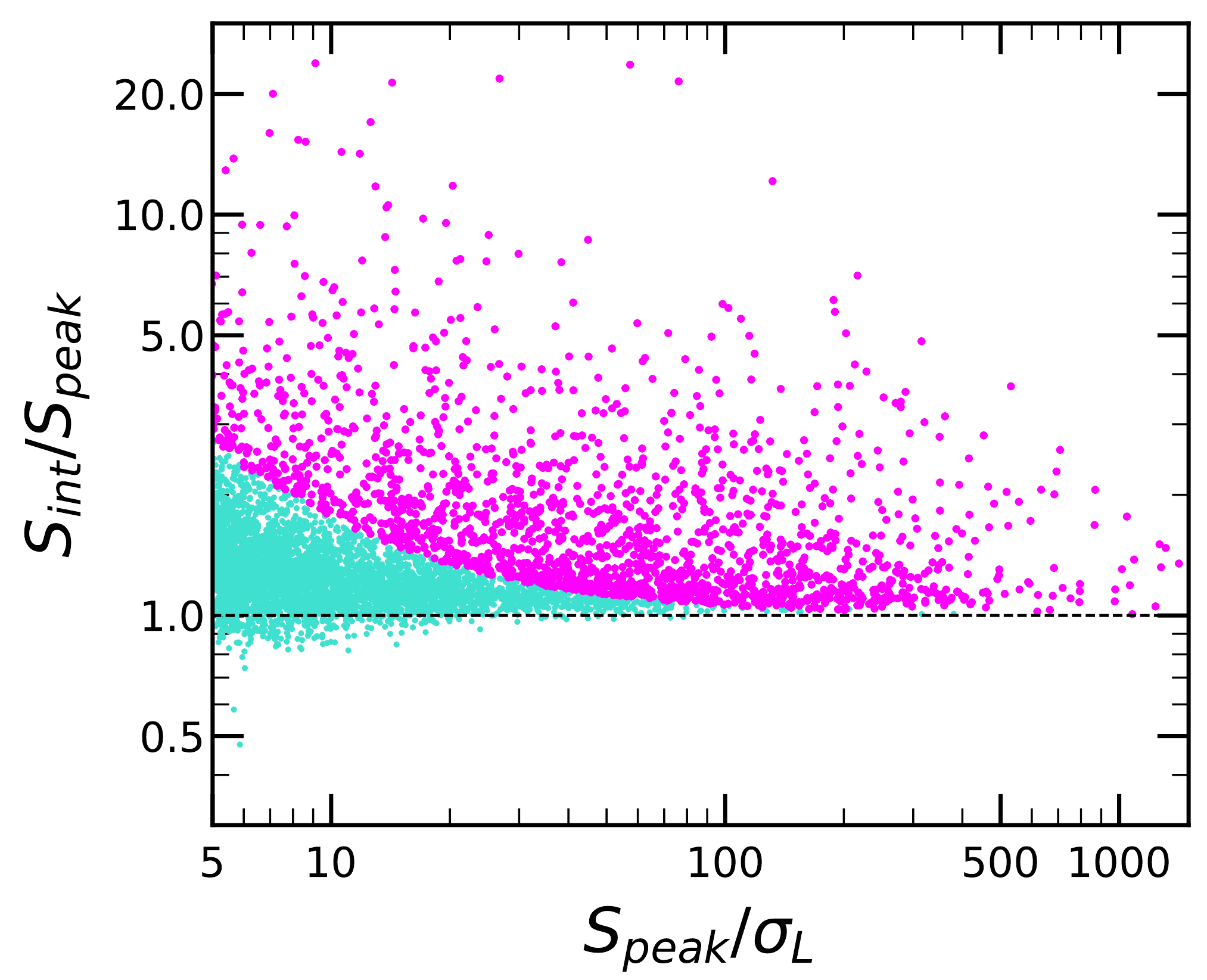}
\caption{Ratio of of integrated-to-peak flux as a function of the source SNR. The teal colored dots are unresolved sources and the magenta filled dots represent resolved sources}
\label{resolved}
\end{figure}

It is not very straightforward to classify the sources detected into resolved and unresolved (point) categories by just using the derived source properties. Time and bandwidth smearing might extend the sources artificially in the image plane. Also, calibration errors, if any, as well as variable noise, may scatter the ratio of integrated flux density ($S_{\mathrm{int}}$) to peak flux density ($S_{\mathrm{peak}}$). Consequently, sources cannot be classified into point and resolved by mere use of $\Big(S_{\mathrm{int}}/S_{\mathrm{peak}}\Big) > 1$. In Figure \ref{resolved}, $\Big(S_{\mathrm{int}}/S_{\mathrm{peak}}\Big)$ as a function of $\Big(S_{\mathrm{peak}}/\sigma_L\Big)$ is plotted, where $\sigma_L$ is the local RMS.

For point sources, assuming that
$\sigma_{\mathrm{S_{peak}}}$ and $\sigma_{\mathrm{S}}$ are independent (where $\sigma_{\mathrm{S_{peak}}}$ and $\sigma_{\mathrm{S}}$ are the uncertainties in the peak flux density and integrated flux density respectively), $ln\Big(S/S_{peak}\Big)$ follows a zero mean Gaussian distribution, with RMS given by 
\begin{equation}
    \sigma_{\mathrm{R}} = \sqrt{\Big(\frac{\sigma_{\mathrm{S}}}{S_{\mathrm{int}}}\Big)^2 + \Big(\frac{\sigma_{S_{\mathrm{peak}}}}{S_{\mathrm{peak}}}\Big)^2}
\end{equation}
An extended source is detected at the 3$\sigma$ level iff R \textgreater 3$\sigma_R$ \citep{franzen2015}. Using this criterion, a total of 1825 sources are resolved (magenta dots), while the remaining 4460 sources are point sources (turquoise dots) as shown in Fig \ref{resolved}.

\section{Comparison with other radio catalogs}
\label{sec.comparison}
This section describes the cross-validation of this work with other radio catalogs covering the same region.  \citet{Garn08} have studied the same area of the sky using 610 MHz observation of the legacy GMRT. There are also several other studies covering various parts of the Lockman Hole region at radio frequencies. Amongst those, the studies by \citet{mahony2016} using 150 MHz LOFAR data and by \citet{prandoni2018} using 1.4 GHz Westerbork Synthesis Radio Telescope (WSRT) data have been considered for comparison. The large area VLA Faint Images of the Radio Sky at Twenty-Centimeters survey (FIRST survey) \citep{Becker1995} has also been used for validation. 

The cross-validation of the source catalog with previous findings is necessary since the ionospheric fluctuations are significant at low-frequency. This will distort the source position and smear the sources residing at a large distance from the phase center. Thus comparison with existing literature allows us for quantification of any systematic offsets in flux densities as well as in source positions. Using a 5$\arcsec$ search radius in the other catalogs, counterparts of the sources presented in this catalog have been identified. Each catalog has a flux density limit depending on the observation sensitivity and completeness. The flux limit of a given catalog was scaled to 325 MHz using $S_{\nu}\propto \nu^{\alpha}$ , where $\alpha$ = -0.8 and  denotes that limit at 325 MHz as $S_{\mathrm{cut,325MHz}}$. Only those sources with flux densities higher than this flux cut-off were chosen. Table \ref{spectral_table} enlists the resolution of each of the chosen catalogs, their corresponding flux limits, and the equivalent 325 MHz cut-off.

\begin{table} 
\begin{center}
\caption{Flux limits of the catalogues considered}
\label{spectral_table}
\begin{tabular}[\columnwidth]{lcccc}
\hline
\hline
catalog & Frequency  & Resolution & $S_{\mathrm{limit}}^{\dagger}$ &  $S_{\mathrm{cut,325MHz}}$ \\
       &  (MHz)   & (arcsec) & (mJy)& (mJy)\\
 \hline
GMRT & 325  & 9.0$\arcsec$ & 0.25 & 0.25 \\
(this work) & & & & \\
\hline
FIRST & 1400  & 5.4$\arcsec$ & 1.0 & 2.779 \\
\citep{Becker1995} & & & & \\
\hline
LOFAR  & 150 & 14.7$\arcsec$ & 2.000 & 1.164\\
\citep{mahony2016} & & & & \\
\hline
GMRT & 610 & 6.0$\arcsec$ & 0.556 & 0.864\\
\citep{Garn08} & & & & \\
\hline
 WSRT & 1400  & 11.0$\arcsec$ & 0.070 & 0.194\\
 \citep{prandoni2018} & & & & \\
\hline
\hline
\end{tabular}
\end{center}
\flushleft{${\dagger}$ $S_{\mathrm{limit}}$ is the flux density limit of the corresponding catalog. \\
 }
\end{table}

 \subsection{Flux Density Offset}
 \label{offset}
 
\begin{figure}

\includegraphics[width=\columnwidth, height=3in ]{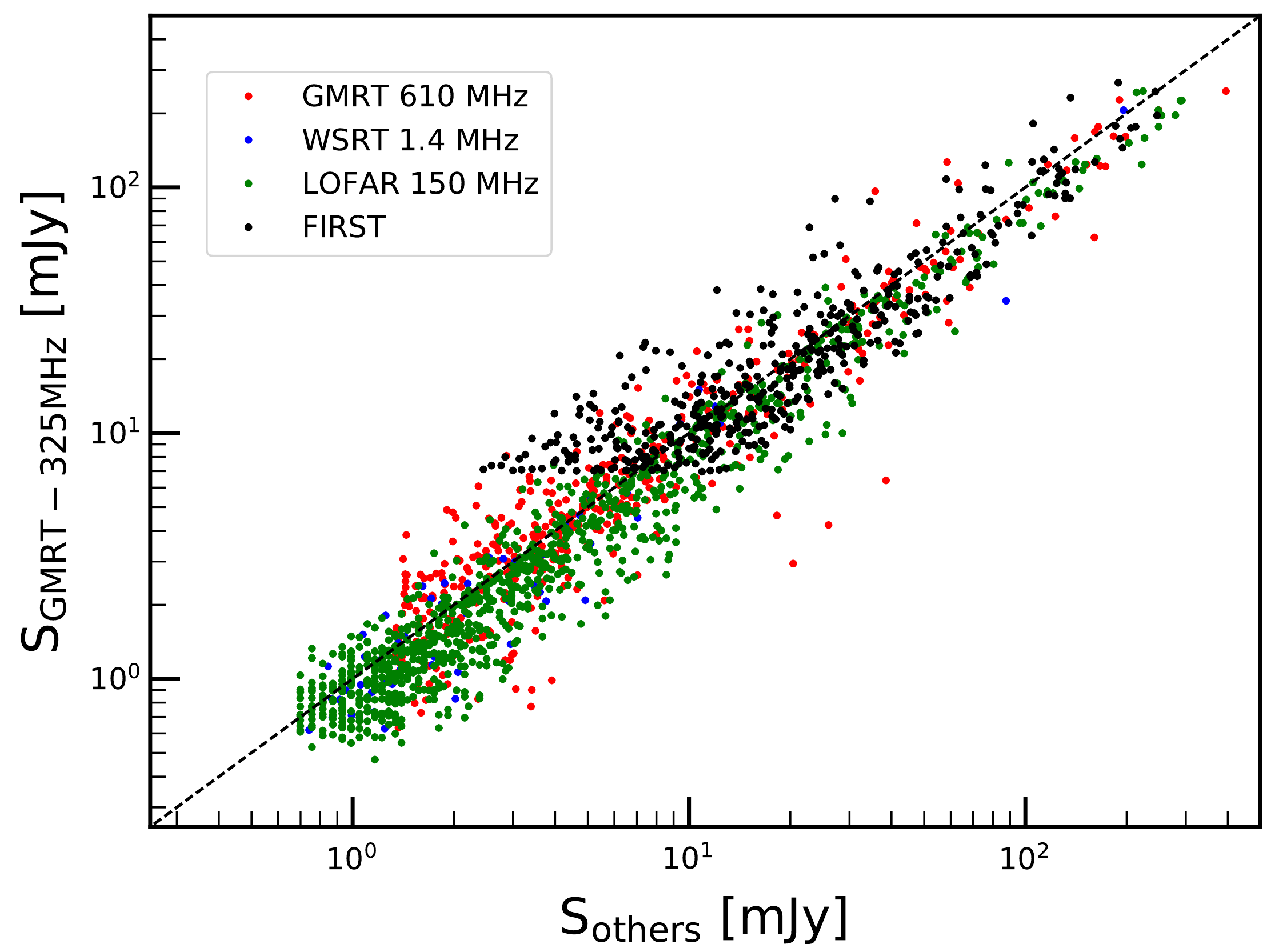}

\caption{Comparison of integrated flux densities of compact sources measured with GMRT at 325MHz with other catalogues: 610MHz GMRT(red), 1.4GHz WSRT(blue), 150MHz LOFAR(green). Fluxes have been scaled using $\alpha$=l-0.8 The black dashed line corresponds to $S_{\mathrm{GMRT}}$/ $S_{\mathrm{others}}$ = 1.  }
\label{flux_ratio_hist}
\end{figure}

 Different catalogs set the flux scales following different flux density scales. Here, the Scaife-Heald flux scale \citep{scafie-heald2012} has been used to set the flux of sources. This scale has also been used in LOFAR 150 MHz data of \citet{mahony2016}. For WSRT 1.4 GHz data of \citet{prandoni2018}, the FIRST catalog \citep{Becker1995} and GMRT 610 MHz observation of \citet{Garn08}, flux standard of \citet{Baars1977} was used. Due to uncertainties in the flux scale used, as well as in modeling of the primary beam, systematic offsets may arise in the flux density of the sources. The flux densities of cataloged sources have been compared with the catalogues mentioned above to check for such systematic offsets. The source selection criterion is based on \citet{williams2016}, where only high SNR sources $\Big(S_{\mathrm{peak}} > 10\sigma\Big)$ have been selected for comparison. Further, the sources are selected to be "compact." This condition implies that the sources have sizes below the resolution at a higher frequency. There was another additional constraint on the flux limit. The catalogues used have a flux limit; for instance in \citet{Garn08} it is $556\mathrm{\mu Jy}$ which corresponds to $920\mathrm{\mu Jy}$ at 325 MHz (assuming $\alpha$ = -0.8). Only sources above such limits (tabulated in Table \ref{spectral_table} for all the catalogues considered) were selected for analysis.
The flux density ratios, after proper scaling of the fluxes, are calculated
\Big($S_{\mathrm{325MHz}}/S_{\mathrm{others}}$, where others mean the previous catalogues used\Big). The median of the ratio comes out to be $0.99^{+0.3}_{-0.5}$ (errors from the 16th and the 84th percentiles) with the 610 MHz data. With FIRST, 150 MHz and 1.4 GHz data, the median values of the flux ratios are  $0.97^{+0.5}_{-0.5}$, $0.82^{+0.3}_{-0.6}$, $0.90^{+0.3}_{-0.6}$ respectively. This is shown in Figure \ref{flux_ratio_hist}, where it is observed that the ratio is nearly 1 for most cases, showing the reliability of the fluxes obtained. 

\begin{figure}
    \centering
    \includegraphics[width=\columnwidth,height=3in ]{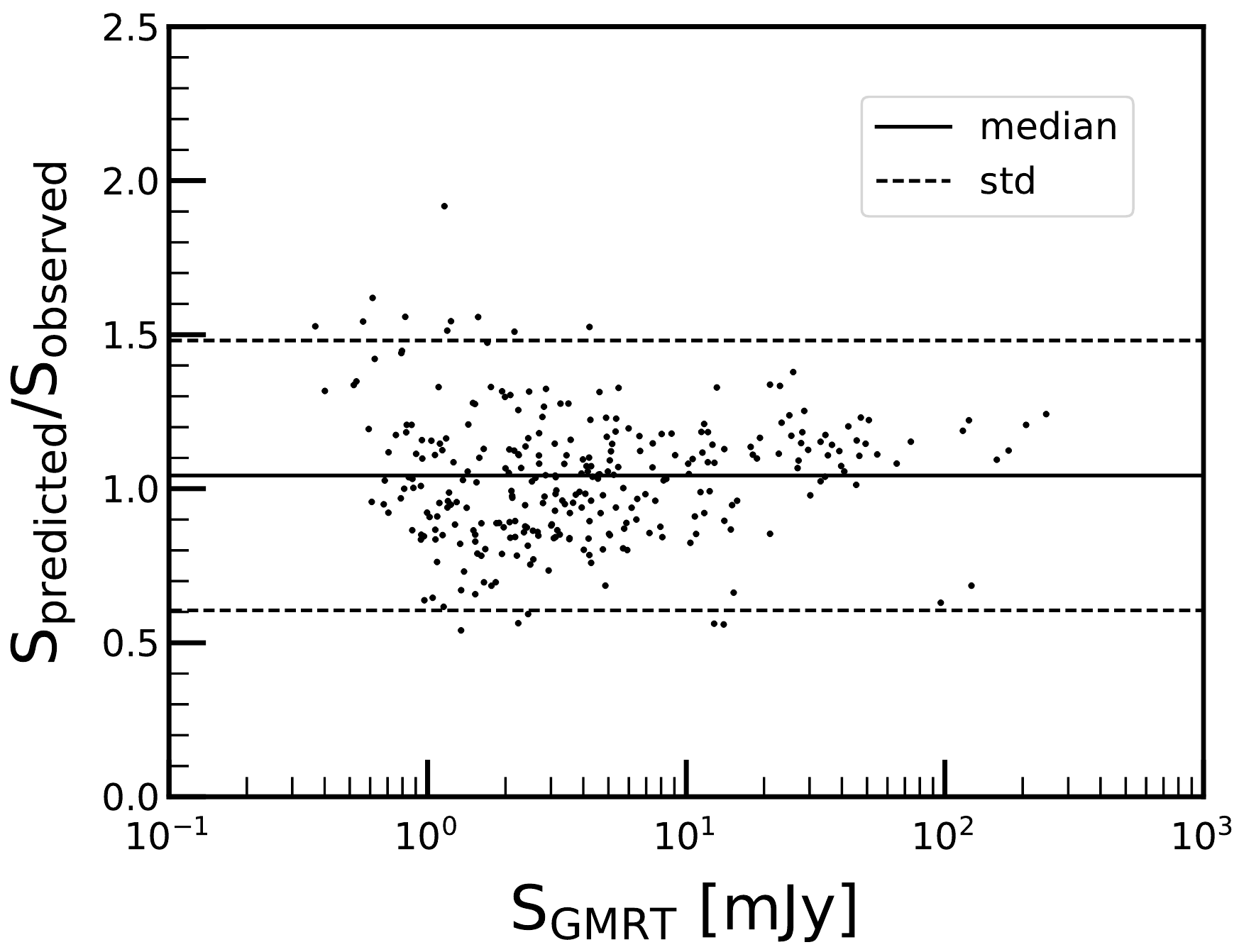}
 \caption{Comparison of flux densities obtained from the 325 MHz data in this work with flux densities predicted using spectral densities obtained from \citet{Garn08} (610 MHz) and \citet{mahony2016} (150 MHz). The solid black line indicates the mean ratio of the predicted and observed flux and the standard deviation of the ratios is indicated by the dashed black lines}
  \label{flux_valid}
\end{figure}

For further validation of the flux scale reliability, sources common to the 610 MHz GMRT catalogue \citep{Garn08} and the 150 MHz LOFAR catalogue \citep{mahony2016} and the catalogue obtained in this work have been selected (satisfying the same criterion described previously). The spectral indices are calculated using the flux densities obtained from the other two catalogues. The mean value for the spectral index is 0.83. Using this value of the spectral index, the fluxes are calculated for the 325 MHz catalogue. Comparison with the actual flux values obtained in the catalogue with the predictions obtained using the value of spectral index gives flux ratios with a median value 1.04 and a standard deviation of 0.43. This has been illustrated in Figure \ref{flux_valid}. As it is clearly seen in the figure, the flux scale used for obtaining source fluxes for this work is reasonably accurate. 

\subsection{Positional Accuracy} 

The astrometric accuracy of the source positions obtained is verified by comparing the source positions of selected sample to the FIRST, LOFAR 150 MHz catalog and GMRT 610 MHz catalog. The source selection criteria also remain the same as previously mentioned in  \ref{offset}. FIRST, being at higher frequency (1.4 GHz), has a resolution 5.4$\arcsec$ and faces lesser ionospheric fluctuation. All such reasons combined make the position accuracy of FIRST better than 1$\arcsec$ \citep{Becker1995}. The offsets have been set as (following \citealt{williams2016}) :
 \begin{eqnarray}
 \delta_{\mathrm{RA}} &=& \mathrm{RA}_{\mathrm{GMRT_{325}}} - \mathrm{RA}_{\mathrm{FIRST}} \\  
 \delta_{\mathrm{DEC}} &=& \mathrm{DEC}_{\mathrm{GMRT_{325}}} - \mathrm{DEC}_{\mathrm{FIRST}} 
 \nonumber
 \end{eqnarray}
 
 The value of the median offsets with respective errors have been shown in Table \ref{offset_table}. 

\begin{table}
\begin{center}
\caption{Median values of the deviation (along with the 16th and 85th percentile errors) of RA and DEC of GMRT 325MHz source catalogue from other catalogues}
\label{offset_table}
\begin{tabular}[\columnwidth]{lccc}
\hline
\hline
catalog & Frequency & $\delta_\mathrm{{RA,median}}$ &  $\delta_\mathrm{{DEC,median}}$ \\
       & MHz & (arcsec)  & (arcsec)  \\
\hline
FIRST & 1400 & $-0.244^{+1.2}_{-1.0}$ & $0.600^{+0.8}_{-0.9}$ \\
\hline
GMRT & 610 & $-0.063^{+0.9}_{-0.8}$ & $0.562^{+0.8}_{-0.6}$ \\
\hline
\hline
\end{tabular}
\end{center}
\end{table}

\begin{figure}

\includegraphics[width=\columnwidth, height=3in ]{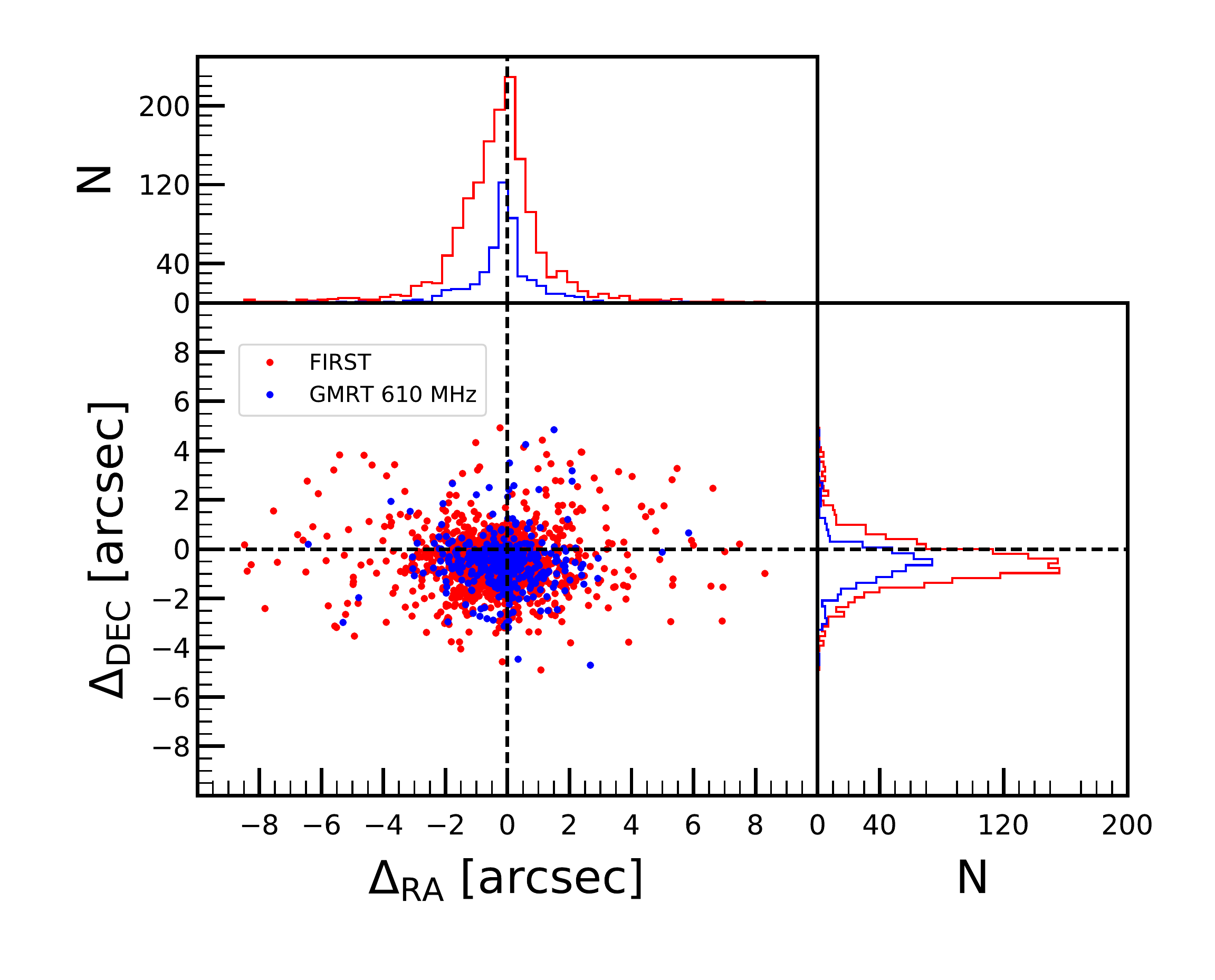}

\caption{Offset of the source RA and Dec for the 325 MHz catalog from FIRST catalog (red) and GMRT 610 MHz catalog (blue).}
\label{ra_dec_hist}
\end{figure}

Figure \ref{ra_dec_hist} shows the histogram of offset values from the other observations considered. No systematic deviation across the field of view has been observed. 
However, the offsets being lesser than the 2$\arcsec$ (i.e., the cell size of the image), there is no significant astrometric error that may be caused by it. The final catalog source positions have been corrected by a constant value derived from the offset from FIRST (the median offset value).


 \subsection{Spectral Index Distribution}
\begin{figure}
    \centering
    \includegraphics[width=\columnwidth,height=2.5in ]{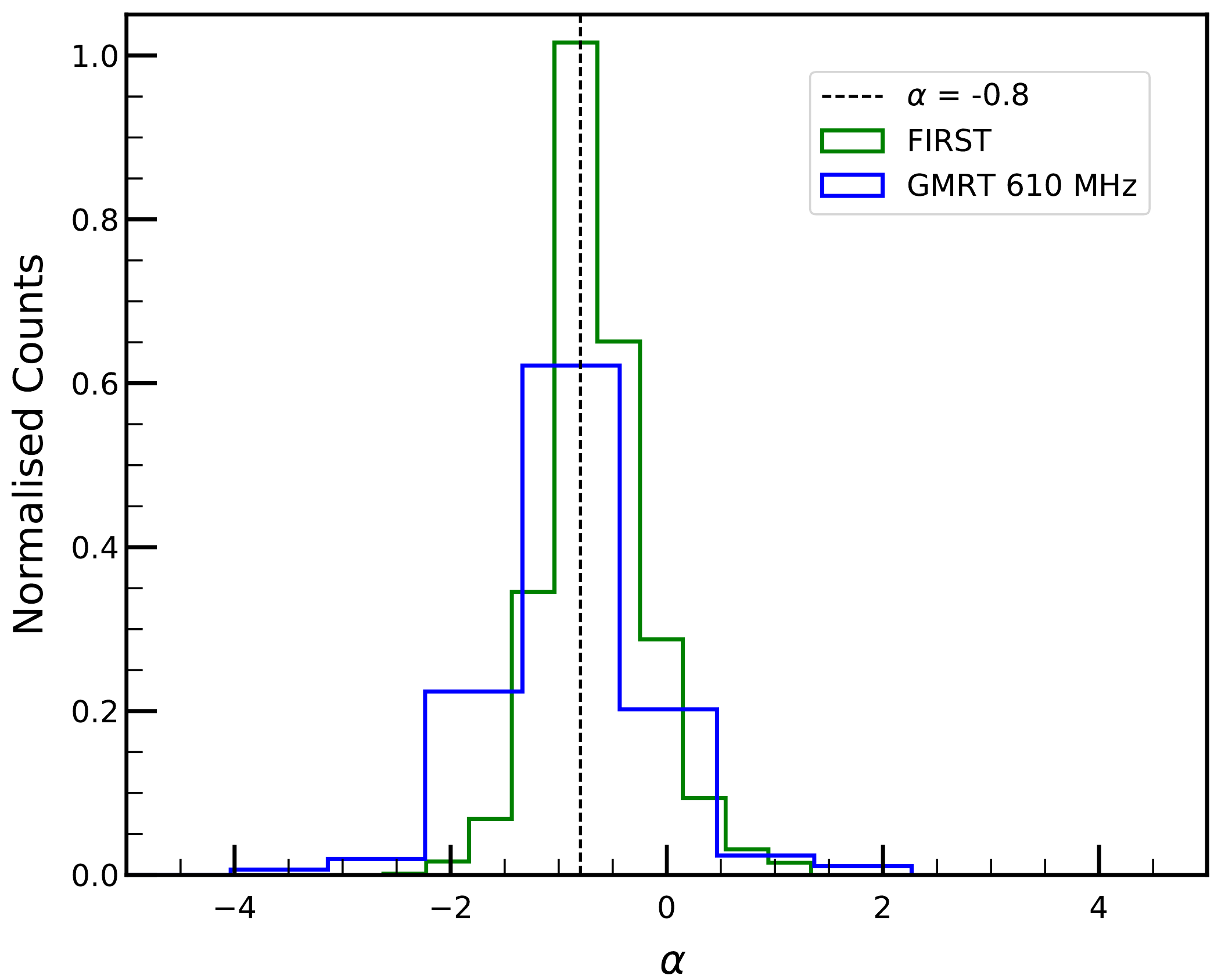}
    \caption{Normalised counts of the measured spectral indices in the field, obtained after matching using a 5$\arcsec$ match radius with FIRST(green) and GMRT 610 MHz data (blue). Black dashed line corresponds to spectral index -0.8}
    \label{alph_hist}
\end{figure}

The present data covers a wide area in the Lockman Hole region with a large number as well as a variety of sources. The spectral properties of the sources are characterized by comparing the fluxes derived for this work with other available data at higher frequencies. The data used are 610 MHz data of \citet{Garn08} and 1.4 GHz FIRST catalog. The source selection procedure is the same as that in section \ref{offset}. The number of sources used to obtain the spectral indices is 511 for GMRT data and 1696 for FIRST. Assuming a synchrotron like power-law distribution, $S_{\nu} \propto \nu^{\alpha}$, where $\alpha$ is the spectral index, the flux densities of the matched sources were used to estimate the value of $\alpha$. The distribution of spectral indices of matched sources is depicted in Figure \ref{alph_hist}. For this work, as is seen in figure \ref{alph_hist}, the normalized counts are highest between the range $\sim$ -1.3 to $\sim$ -0.4. 
The median values of the spectral indices with errors from the 16th and 84th percentile are $-0.860^{+0.621}_{-0.578}$, $-0.708^{+0.344}_{-0.506}$ for GMRT 610 MHz and FIRST catalogues respectively. Hence, the median value for the spectral index of this catalog after matching can be taken as $\sim$ -0.8.
Several low frequency radio observations for the Lockman Hole region has spectral index distribution with $\mid\alpha\mid \leq$ 2.0 (for instance \citealt{Garn08,ibar09,mahony2016}). Low frequency studies for other deep fields like COSMOS \citep{smolcic2017}, Bo\"{o}tes \citep{williams2016}, ELAIS-N1 \citep{arnab2, ocran} also mostly follow this distribution (except a few cases seen in \citealt{smolcic2017} (Figure 14) \& \citealt{ocran} (Figure 14)). But it can be seen in Figure \ref{alph_hist} that the distribution of $\alpha$ is wider than what is usually observed. The exact reason for this is unknown. However, since a detailed study of  spectral index requires multi-wavelength as well as wider bandwidth data, it is deferred to future works.

\section{Source counts}
\label{sec.dnds}

This section presents the differential source counts based on the flux densities arising in the P{\tiny Y}BDSF generated outputs. At low frequencies, the most dominant source population at the faint flux end is that of SFGs and Radio-quiet Quasars (RQQs). This has been well established from various observations as well as simulations (\citealt{trecs, S3}, and references therein). Nevertheless, observational constraints on sub-mJy source populations are very few. However, for telescopes like MWA, SKA, or LOFAR, which aim to detect the faint cosmological HI 21-cm signal, the sources from sub-mJy flux levels down to$\mu$Jy level act as foregrounds which can potentially obscure the signals. Hence, characterizing the spatial and spectral nature of sources with flux densities from sub-mJy to $\mu$Jy, especially at low frequencies, is essential for foreground characterization. The determination of source counts, i.e., the source distribution at the flux ranges of interest, is one of the significant steps for such characterization.

The differential source counts at 325 MHz have been measured with sources having flux densities down to 0.2mJy ($\approx$ 4$\sigma$). However, this does not give the correct distribution, since the  P{\tiny Y}BDSF output has errors arising out of catalog incompleteness, resolution bias, false detection, Eddington bias. The problem is especially prevalent at low frequencies and for the faint end of the flux bins. The following subsections describe the corrections made to the source count distribution in detail.

\subsection{False Detection Rate}

False Detection Rate (FDR) is simply the number of spurious detection by the source finder. This occurs either due to noise spikes or due to the presence of bright artifacts in the image. If noise distribution is symmetric about the mean, i.e., positive noise spikes have equivalent negative peaks in the image, the number of spurious detection would be equal to the number of negative sources in the inverted image. 

To quantify this false detection rate, P{\tiny Y}BDSF was run on the inverted image with parameters identical to that used in the original image. This yielded a total of 71 sources with negative peaks below -5$\sigma$. Now, for FDR correction to the flux density bins, the negative sources are binned in the same manner as the actual sources and compared to the sources detected in the original image. The number of real sources in each bin (quantified as a fraction) is \citep{Hale19}: 
\begin{equation}
     \textit{f}_{\mathrm{real},\textit{i}} = \frac{\textit{N}_{\mathrm{catalog},\textit{i}} - \textit{N}_{\mathrm{inv},\textit{i}}}{\textit{N}_{\mathrm{catalog},\textit{i}}},
    \label{FDR_eqn}
\end{equation}

where  $\textit{N}_{\mathrm{inv},\textit{i}}$ and $\textit{N}_{\mathrm{catalog},\textit{i}}$ are the number of  detected sources in $i^{th}$ flux density bin for inverted and original image respectively.
The errors are quantified as Poisson errors; the correction factor is multiplied with the number of sources in each flux density bin of the original catalogue. The correction factors obtained for each bin is shown in Figure \ref{fdr_comp_fig}.

\subsection{Completeness}

The incompleteness of a catalog is the inability to detect sources lying above the flux limit of the catalog. This is mainly due to noise variation in the image. The source catalog constructed using the source finder algorithm (P{\tiny Y}BDSF) is completeness limited. Incompleteness can cause both underestimation as well as overestimation of the source counts derived from the image plane. Eddington bias \citep{Eddington} causes scattering of higher count bins into lower count ones more than the reverse, subsequently overestimating the latter. Resolution bias is another factor limiting catalog completeness. It arises because the detection probability is lesser for resolved sources than unresolved ones, resulting in an underestimation of source counts.


Simulations were performed on the image plane to quantify and correct for the above biases. 3000 sources were artificially injected in the residual RMS map (same as \citealt{williams2016}) using the open-source software Aegean\footnote{\url{https://github.com/PaulHancock/Aegean}} \citep{hancock1, hancock2}. Of these, 1000 are extended sources (major and minor axes greater than 9$\arcsec$) while the rest are unresolved point sources. This number distribution is done following the actual source classification in the catalog, where $\sim$30\% are resolved. The flux density is generated using  power-law distribution of the form dN/dS $\propto$ $S^{-1.6}$ \citep{intema2011, william2013}. The flux values are chosen randomly and constrained between 200$\mu$Jy to 4100$\mu$Jy. 
The source positions were also randomly chosen, spanning the entire right ascension and declination range of the catalog. Following \citet{arnab2}, 100 different realizations of the said simulations were done. The simulations take into account the visibility area effects and source confusion limitations \citep{Hale19,franzen2019, williams2016}.     

The sources are extracted separately from each image using P{\tiny Y}BDSF, setting the same parameters as described in Section \ref{sec.source_catalog}. The recovered sources are binned into the same number of bins as the actual catalog. The correction factor is calculated as:

\begin{equation}
     \mathrm{Correction}_{,\textit{i}} = \frac{\textit{N}_{\mathrm{injected},\textit{i}}}{\textit{N}_{\mathrm{recovered},\textit{i}}}
\end{equation}

where, $\mathrm{Correction}_{,\textit{i}}$ is the completeness correction factor in the $i^{th}$ flux density bin $\textit{N}_{\mathrm{injected},\textit{i}}$ is the number of injected sources and  $ \textit{N}_{\mathrm{recovered},\textit{i}}$ is the number of sources recovered after subtracting original pre-simulation sources in the $i^{th}$ bin \citep{Hale19}.

\begin{figure}
    \centering
    \includegraphics[width=\columnwidth,height=2.5in ]{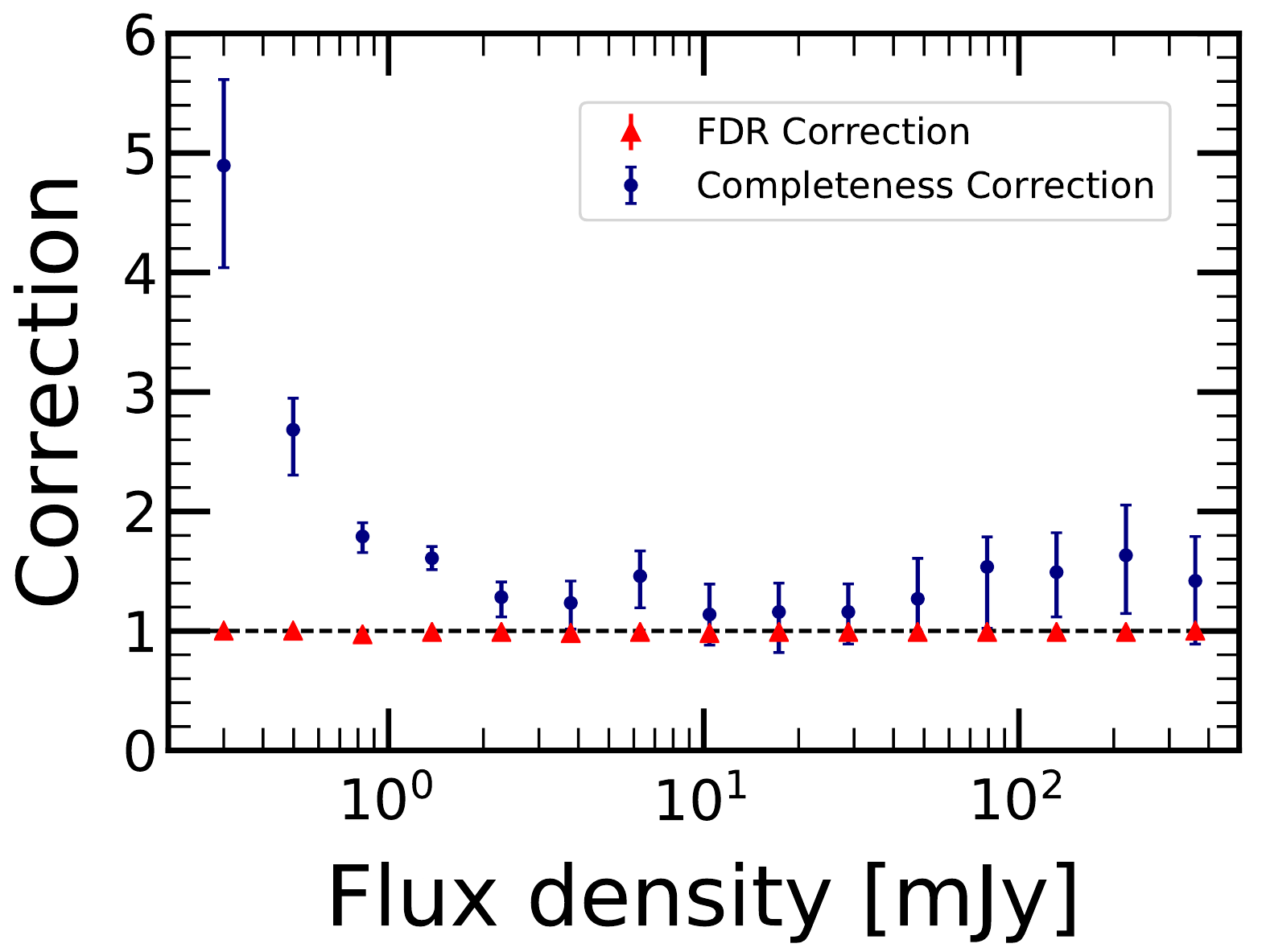}
    \caption{Correction factors for false detection (red triangles) and completeness (blue circles).}
    \label{fdr_comp_fig}
\end{figure}

Figure \ref{fdr_comp_fig} show the FDR and completeness correction factors for each of the flux bins. The median value of each flux bin from the 100 simulations is taken as the correction factor, while the errors are those associated with the 16th and 85th percentiles.

 \subsection{Differential Source Count}
 
\begin{figure*}
    \centering
    \includegraphics[width=6.0in, height=4.0in]{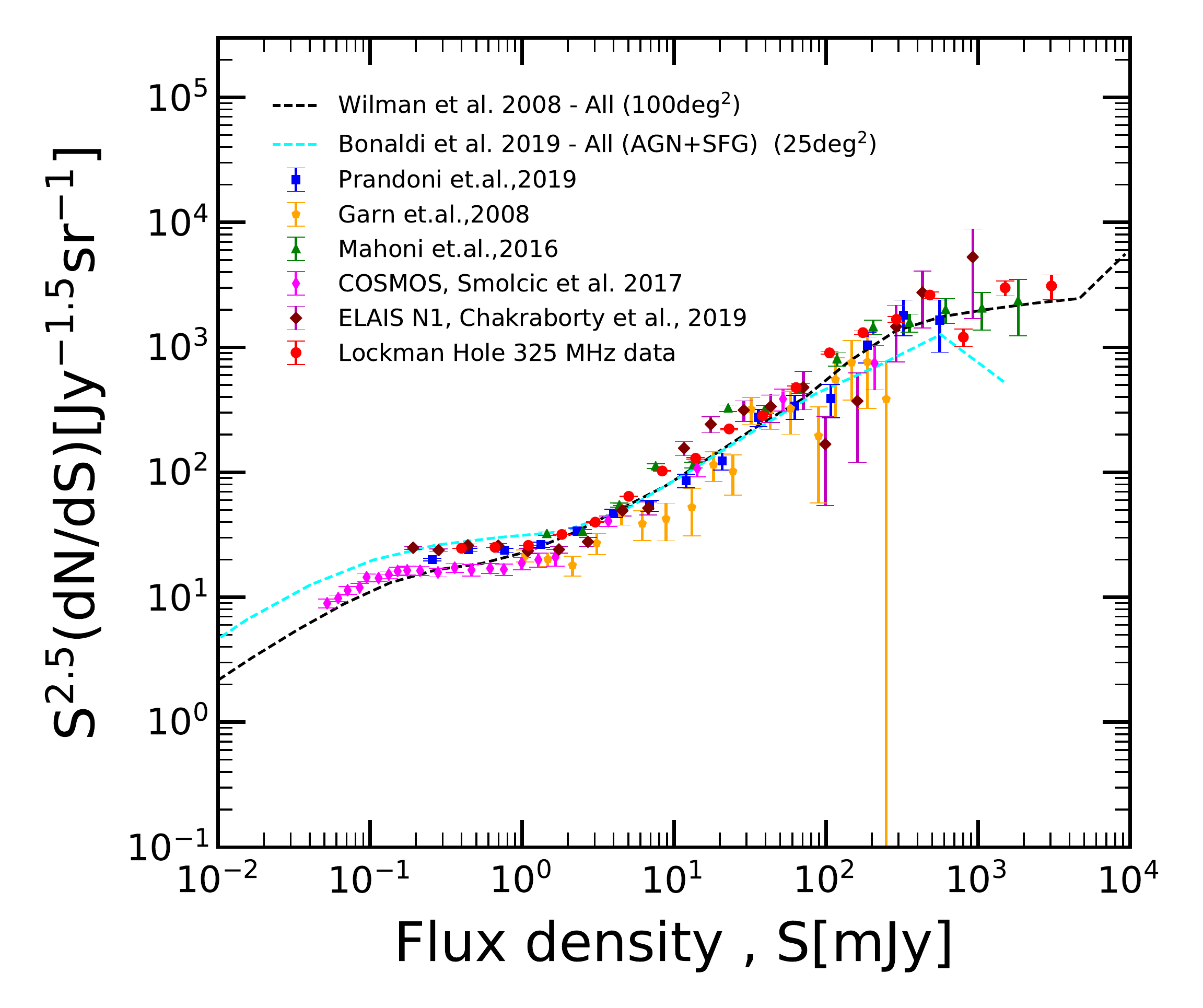}
    \caption{Euclidean normalised source counts for the 325 MHz data (shown in red circles) after correcting for false detection and incompleteness . It has been compared against source counts modelled by simulations - S$^3$ \citep{S3} plotted as black dashed curve and T-RECS \citep{trecs} plotted as cyan dashed curve. Comparison has also been shown with other observation covering the same region of the sky - WSRT 1.4 GHz \citep{prandoni2018} (blue squares), GMRT 610 MHz \citep{Garn08} (orange pentagons) \& LOFAR 150 MHz \citep{mahony2016} (green triangles). Comparison has also been done with source counts for VLA COSMOS 3 GHz survey \citep{smolcic2017} (magenta diamonds) \& GMRT 400 MHz observation of ELAIS N1 \citep{arnab2}(maroon diamonds)}
    \label{dnds}
\end{figure*}

The Euclidean-normalised differential source counts have been estimated from the P{\tiny Y}BDSF generated source list, after correcting for incompleteness and FDR. The correction factor for each bin is multiplied to the respective uncorrected source counts. Since noise is variable across the image (see Figure \ref{rms}), correction is also done for effective area per bin (i.e., area over which a source can be detected). To determine the effective area per bin, the fraction(f) of the area over which a source having a specific flux density is detectable has been determined. The source count in each bin is then weighted by f$^{-1}$ \citep{Windhorst1985}. The fluxes have been binned into 18 logarithmic bins down to 0.30mJy (6$\sigma$), and Poisson errors on the source counts have been evaluated.  
The source counts and associated errors are given in Table \ref{table_source_count}. Post incorporation of the required corrections, the normalized differential source counts have been plotted in Figure \ref{dnds}. Comparison has also been made with state of the art simulations as well as other observed counts. 

The differential source counts have been measured against the source counts of 610 MHz GMRT data of the Lockman Hole region \citep{Garn08}, after scaling the fluxes to 325 MHz using a spectral index of -0.8. Using the same spectral index, the counts have also been compared against differential source counts at 1.4 GHz \citep{prandoni2018} and those at 150 MHz \citep{mahony2016}. The same comparison was also done with   $S^{3}$ -SKADS simulation \citep{S3} and T-RECS simulation\citep{trecs}. The fluxes at 1.4 GHz from both the catalogs were scaled using a spectral index -0.8.  For simulating the $S^{3}$ -SKADS catalog, sources are drawn from luminosity functions determined via various multi-frequency observations, and implanted into an underlying dark matter density field, with bias factors attributed to the observed clustering at large scales (see \citealt{S3} for details). In T-RECS, updated evolutionary models were used to separately model two main populations of radio galaxies: AGNs and SFGs with their respective sub-populations. The obtained luminosity functions were validated using available data and were found in agreement (see \citealt{trecs} for further details).  For further validation, the fluxes have also been matched with other observations - GMRT observation of the ELAIS N1 field at 400 MHz \citep{arnab2} and 3 GHz VLA-COSMOS survey \citep{Smolic08}. As can be observed in Figure \ref{dnds}, the source counts for this work (red circles), match with the previous surveys as well as simulations. The counts flatten downwards of 1 Jy, which is attributed to the increase in the number of SFGs (which dominate at these flux values), increasing overall source counts. 

\begin{table*}
\caption{Euclidian-normalized differential source counts for the Lockman Hole field.}
\label{table_source_count}
\scalebox{1.0}{
\begin{tabular}{c c c c c c c c}
\hline
\hline
 S & $S_{c}$  & N   & $ S^{2.5}$dN/dS & FDR & Completeness &  Corrected $ S^{2.5}$dN/dS  \\
(mJy) & (mJy) &       &  ($\mathrm{Jy}^{1.5}$sr$^{-1}$)  &       &    & ($\mathrm{Jy}^{1.5}$sr$^{-1}$)           \\
 \hline
 0.3 - 0.5 & 0.4	& 260	& $4.9 \pm 0.1$ &	$1.0 \pm 0.0$ & $4.9_{-0.9}^{+0.7}$	& $24.7 \pm 0.05$ \\
 & & & & & & & \\
0.5-0.8 & 0.7 & 1315 &  $9.3 \pm 0.2$ & $1.0 \pm 0.0$ & $2.7_{-0.4}^{+0.3}$ & $25.1 \pm 0.06$ \\
& & & & & & & \\
0.8 - 1.4 & 1.1 & 1316 &	$14.9 \pm 0.4$ & $0.97 \pm 0.0$ & $1.8_{-0.1}^{+0.1}$ &	$25.9 \pm	0.05$ \\
& & & & & & & \\
1.4 - 2.3 & 1.8 &	906 &  $19.9 \pm 0.7$ & $0.99 \pm 0.0$ & $1.6_{-0.1}^{+0.1}$ &	$31.8	\pm 0.06$ \\
& & & & & & & \\
2.3 - 3.8 & 3.0 & 680 & $31.4 \pm 1.2$ &	$0.99 \pm 0.0$	& $1.3_{-0.2}^{+0.1}$ &	$39.9 \pm 0.15$ \\
& & & & & & & \\
3.8 - 6.3	& 5.0 & 536 & $53.0 \pm 2.3$ &	$0.98 \pm 0.0$ & $1.2_{-0.2}^{+0.2}$ &	$64.2	\pm 0.42$ \\
& & & & & & & \\
6.3 - 10.4 & 8.4 & 337 & $70.9 \pm 3.9$ & $0.99 \pm 0.0$	& $1.5_{-0.3}^{+0.2}$ &	$102.4 \pm	0.81$\\
& & & & & & & \\
10.4 - 17.3 & 13.9 & 258 & $116.1 \pm 7.2$ & $0.98 \pm 0.0$ & $1.1_{-0.3}^{+0.3}$ & $129.5 \pm	1.84$ \\
& & & & & & & \\
17.3 - 28.8	& 23.0 & 201	& $193.6 \pm 13.7$ & $0.99 \pm 0.0$	& $1.2_{-0.3}^{+0.2}$ & $222.1 \pm 3.3$ \\
& & & & & & & \\
28.8 - 47.7 & 38.3 & 120 & $247.2 \pm 22.6$ & $0.99 \pm 0.0$ & $1.2_{-0.3}^{+0.2}$ &	$283.7	\pm 5.3$\\
& & & & & & & \\
47.7 - 79.3 & 63.5 & 86	& $379.0 \pm 40.9$ & $0.99\pm 0.0$ & $1.3_{-0.3}^{+0.3}$ &	$476.2	\pm 13.9$ \\
& & & & & & & \\
79.3 - 131.6 & 105.4 & 63 & $593.9 \pm 74.8$ & $0.99 \pm 0.0$	& $1.5_{-0.5}^{+0.3}$	& $903.2 \pm 18.8$ \\
& & & & & & & \\
131.6 - 218.5 & 175.1  & 44 & $887.4 \pm 133.8$ & $0.99 \pm 0.0$ & $0.8_{-0.2}^{+0.2}$ & 1310.8 $ \pm 44.2$ \\
& & & & & & & \\
218.5 - 362.8 & 290.6 & 23 & $1035.5 \pm 211.4$ & $0.99 \pm 0.0$ & $1.5_{-0.4}^{+0.3}$ &	$1674.2 \pm 88.8$ \\
& & & & & & & \\
362.8 - 602.3 & 482.5 & 20 & $1846.1 \pm 412.8$ & $1.00 \pm 0.0$ & $1.4_{-0.5}^{+0.4}$& $2619.6	\pm 153.2$ \\
& & & & & & & \\
602.3 - 1000.0 & 801.2 & 4 & $789.9 \pm 394.9$ & $0.99 \pm	0.0$ & $1.5_{-0.4}^{+0.5}$ &	$1205.8 \pm 190.8$ \\
& & & & & & & \\
1000.0 - 2018.5 & 1509.3 & 5 & $1877.9 \pm 839.8$ & $0.97 \pm 0.0$ & $1.6_{-0.9}^{+0.5}$ & $2992.7 \pm 410.7$ \\
& & & & & & & \\
2018.5 - 4074.3 & 3046.4 & 2 & $2154.1 \pm 1523.2$ & $0.99 \pm 0.0$ & $1.5_{-0.9}^{+0.5}$ &	$3098.6 \pm 689.9$ \\
& & & & & & & \\
 \hline 
\end{tabular}}
\flushleft{Notes: This table includes the flux density bins, central of flux density bin, the raw counts, normalized source counts, False Detection Rate (FDR), completeness and corrected normalized source counts.}
\end{table*}

\section{Power Spectrum of Diffuse Emission}
\label{sec.DGSE}

Diffuse Galactic Synchrotron Emission (DGSE) is one of the major contributing factors to the foregrounds, which obscure the cosmological HI signal from reionization and cosmic dawn. It remains dominant even after modeling and removal of the point sources from the data-sets. However, the spectrally smooth nature of the emission may be exploited for isolating it from the data to extract the faint cosmological target signal. Therefore, detailed knowledge of the spectral as well as spatial characteristics is essential for reasonable removal of foregrounds (i.e., removing it without removing the signal). 

The spectrally smooth nature of the DGSE is captured by modeling it as a power law in both frequency and angular scales (Equation \ref{power_law_model}).
Observational measurements of the angular power spectrum of DGSE for various fields have constrained its power-law index in the range [1.5 to 3.0] \citep{ali08, bernardi09, icobelli13, tge17, tge2020}. Another measurement of the angular power spectrum of galactic synchrotron was done by \citet{laporta08} using the 408 MHz Haslam map \citep{Haslam1982} and  1.4 GHz map of \citet{Reich1988}. The spectral index obtained from the aforementioned sources (which fall in the range [2.9 to 3.2] at different galactic latitudes), were used to extrapolate up-to 23 GHz using a single power-law fit. However, in \citet{arnab2}, using wideband GMRT data, they have shown that the spectral nature of the DGSE has a hint of a broken power law.

The angular power spectrum for the field was quantified using TGE, \citep{tge14, tge16}. It estimates the angular power spectrum using the correlation between grided visibilities. Bright sources near the edge of the FoV can make the spectrally smooth diffuse foreground oscillate, thereby making it challenging to remove. Also, if side lobes of these sources are near the nulls of the primary beam of the telescope, complications arise in extracting the signal. TGE handles this by tapering the primary beam much before the first null is reached. It is also computationally efficient since it grids the visibilities before computing the power spectrum. 

The estimator ($\mathrm{\hat{E}_g}$) is defined as : 
\begin{equation}
\mathrm{\hat{E}_g} = M_g^{-1}\Big(\arrowvert\mathcal{V}_{cg}\arrowvert^{2}-\Sigma_i\arrowvert\Tilde{\omega}{_g}(\mathbf{U_g - U_i})\arrowvert^{2}\arrowvert\mathcal{V}_{i}\arrowvert^{2} \Big)
\end{equation}

where $\mathcal{V}_{cg}$ is the convolved visibility at every grid point $g$, $\mathcal{V}_{i}$ is the measured visibility, $\Tilde{\omega}{_g}$ is the Fourier transform of the window function used for tapering the sky response $\mathbf{U_g}$ refers to baseline corresponding to the grid point $g$ and $M_g^{-1}$ is a normalization factor (refer to \citealt{tge16} for further details). The tapering function used is a Gaussian window function of the form $\mathcal{W}(\theta)=\mathrm{exp(-\frac{\theta^{2}}{\theta_{w}})}$, where $\theta_{w}=\mathrm{f}\theta_{0}$ ($\theta_{0}=\mathrm{0.6} \times\theta_{FWHM}$ of the telescope primary beam).

\begin{figure*}

\includegraphics[width=\columnwidth, height=3in ]{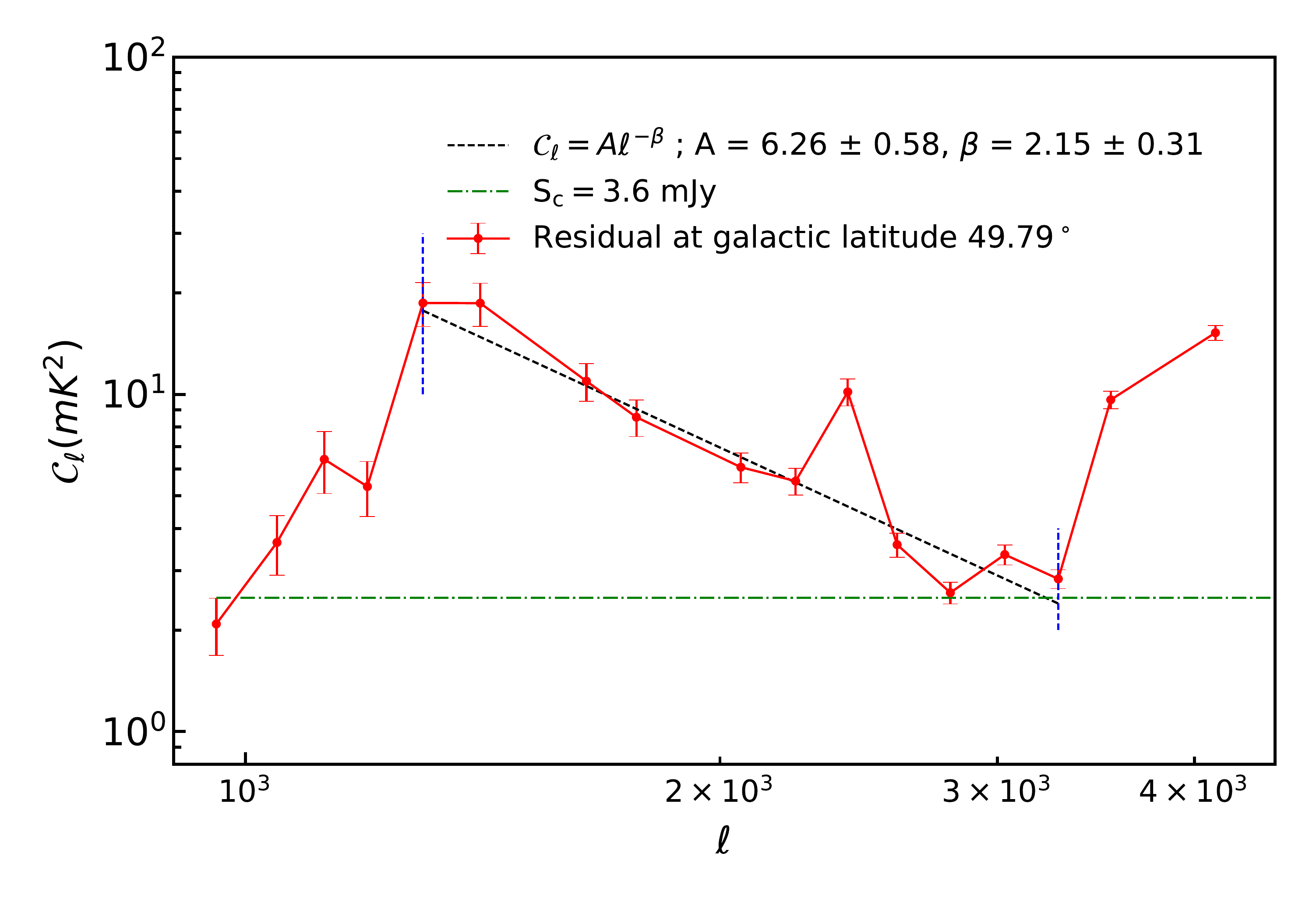}
\includegraphics[width=\columnwidth, height=3in ]{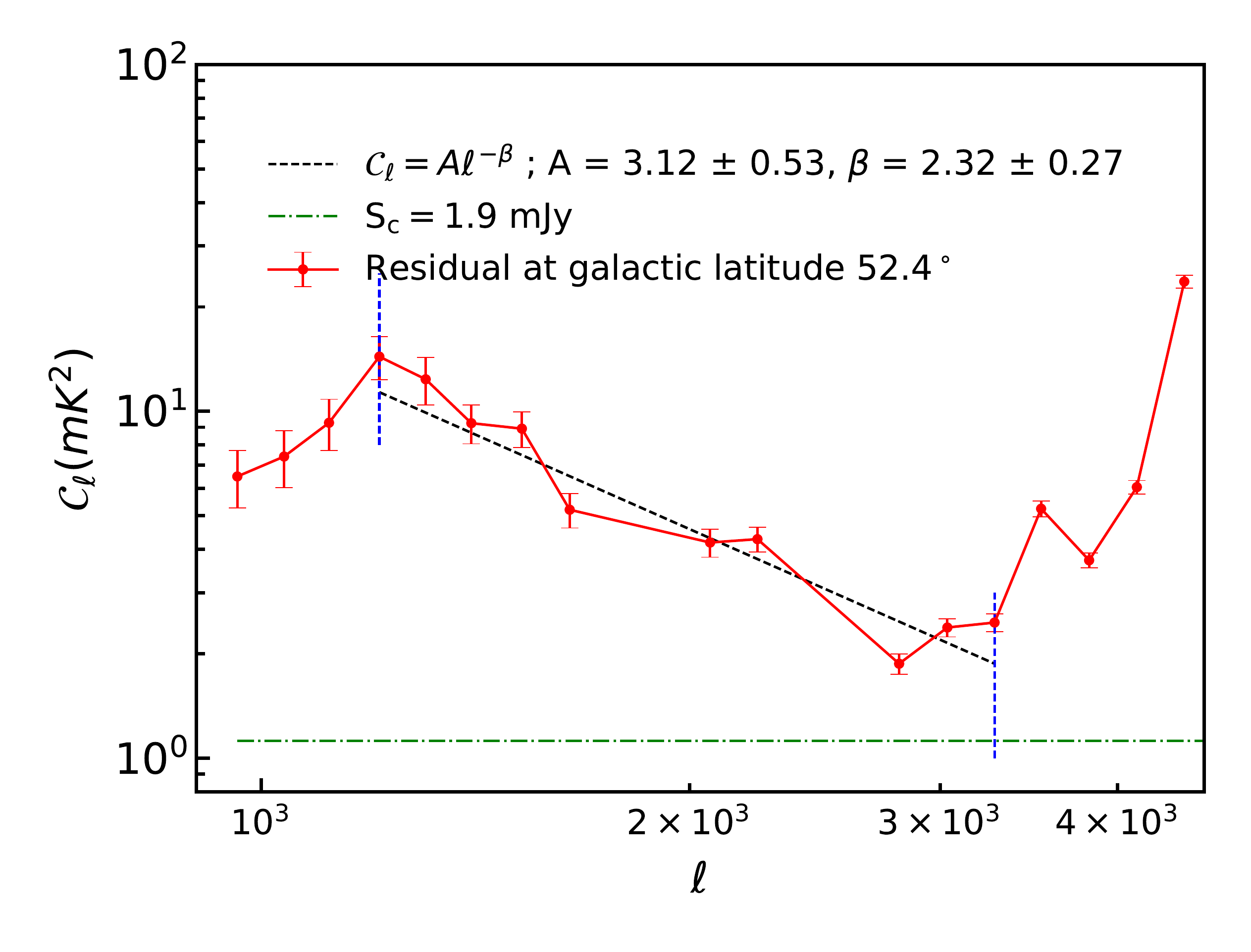}
\includegraphics[width=\columnwidth, height=3in ]{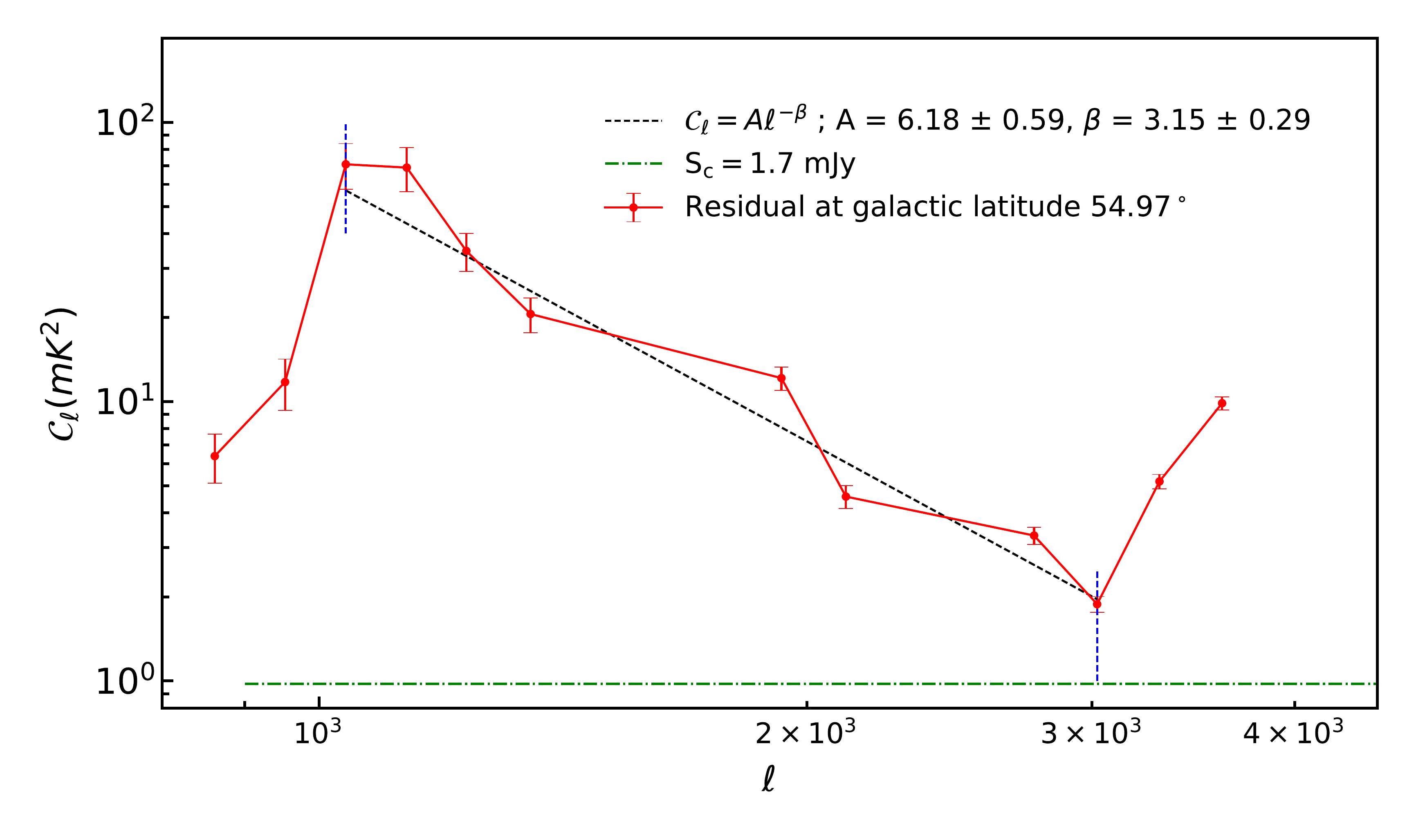}

\caption{Estimated Angular Power Spectrum, $\mathcal{C}_\ell$, with $1-\sigma$ error bars  for 3 different galactic latitudes plotted as a function of angular mode, $\ell$. Red circles represent the observed values and black dashed curve is the best fit power law curve of the form $A\ell^{-\beta}$. The blue dashed lines is the $\ell$ range where the power law could be fit. The green dot-dashed line is the contribution from the unmodeled point sources in the residual data.} 
\label{tge1}
\end{figure*}

The residuals obtained from visibilities calibrated by SPAM should contain only diffuse emission (considering perfect modeling and removal of discrete sources). However, since there are imperfections in the modelling, there will be residual(unsubtracted) point sources with the diffuse emission. To determine the angular power spectrum of the DGSE, TGE was applied to the residual visibilities (obtained after subtracting the point source contribution from the calibrated visibilities). The 23 pointings are spread over a galactic latitude coverage of $\sim$ 3$^\circ$, from which three pointings were chosen at three different galactic latitudes to characterize the variation of DGSE. The values used for determination of $\mathcal{W}(\theta)$ are f=1.0 and $\theta_{0}$=44' for all the three pointings. Figure \ref{tge1} shows the angular power spectrum ($\mathcal{C}_\ell$) as a function of the angular mode $\ell$ plotted for galactic latitudes \textit{b} $\sim$ 50$^\circ$, 52$^\circ$ \& 55$^\circ$. Power-law of the form $A\ell^{-\beta}$ has been fitted to the residual data choosing an $\ell$ range where there is a plunge in the amplitude of power spectrum. The best fit values for A and corresponding power-law index $\beta$ is shown in the plots and also indicated in Table \ref{l_table}. The power-law indices obtained from the best fit values are $\beta$ = 2.15, 2.27 \& 3.15 at \textit{b} = 49.8$^\circ$, 52.4$^\circ$ \& 54.9$^\circ$ respectively. The fall of the $\mathcal{C}_\ell$ values as a function of $\ell$ is consistent with the nature expected for DGSE. The $\ell$ range for fit, along with other details of the pointing, is shown in Table \ref{l_table}. 

The value of U$_{min}$ (corresponding to shortest baseline) for the three pointngs are 50$\lambda$, 80$\lambda$ and 40$\lambda$. These values translate to $\ell_{min}$ 314, 502, and 251, respectively. But these values are ideal; estimating values of $\ell_{min}$ without any bias require consideration for the convolution effects of the primary beam, tapering window, actual uv-coverage (see equation 6 of \citealt{tge16}). The value of $\ell_{min}$ for each field was chosen conservatively from a careful inspection of the obtained power spectrum. In \citet{tge17}, it was shown that at larger angular scales, since convolution effects become important, it may lead to the drop-off in the power seen at $\ell$ below $\ell_{min}$. 

DGSE dominates the power at small angular scales up to a specific $\ell_{max}$.  This value of $\ell_{max}$ was chosen by visual inspection at the $\ell$, where the power-law behavior shows clear breakage. Beyond this point, residual point sources, bright artifacts, and calibration effects may contribute to the obtained power. Beyond the $\ell$ ranges considered,  the power spectrum behavior is the same as seen in earlier studies \citep{tge17}.

Unsubtracted point sources present in the residual can cause Poisson fluctuations in the values of $\mathcal{C}_\ell$. The green dashed curve in Figure \ref{tge1} shows the value of $\mathcal{C}_\ell$ due to presence of residual point sources below a threshold flux (which corresponds to the maximum flux in the residual data) of $\mathrm{S_c}$ = 4.6, 1.9 \& 1.7 mJy respectively from lowest to highest galactic latitudes considered. This is predicted using the formulation described in \citet{ali08}. As seen in Figure \ref{tge1}, the values of $\mathcal{C}_\ell$ where the power-law fitting is obtained lie above the theoretical threshold for residual point source contamination. It is to be noted that the threshold (indicated by green dashed curves) are the lower limits of the expected residual point source contribution under ideal conditions with an assumed dN/dS distribution. However, for the case of real data, considering the presence of residuals, non-Gaussian noise, and calibration errors, a flat floor is not obtained (as seen in Figure \ref{tge1}. Previous studies by \citet{tge17} for TGSS fields have also found similar results.

\begin{table*} 
\begin{center}
\caption{Details of the pointings used for determining the Angular Power Spectrum and the best fit values for A \& $\beta$}
\label{l_table}
\begin{tabular}[\columnwidth]{ccccccc}
\hline
\hline
(RA , Dec) & $(l,b)$ & $\ell_{min}$ & $\ell_{max}$ & A & $\beta$ & $\chi^{2}_{reduced}$ \\
(h:m:s , d.m.s) & (deg,deg)  & &  & &\\
\hline

(10:33:41, 59.44.51) & (149.12, 49.79) & 1200 & 2800 & 6.26$\pm$0.58 & 2.15$\pm$0.31 & 0.89 \\
(10:48:30,58.03.00) & (149.26, 52.37) & 1300 & 3280 & 3.84$\pm$0.22 & 2.27$\pm$0.20 & 0.30\\
(10:55:54, 56.21.09) & (146.14, 54.97) & 1040 & 3020 & 6.18$\pm$0.59 & 3.15$\pm$0.29 & 3.60\\
\hline
\end{tabular}
\end{center}
\end{table*}

The $\mathcal{C}_\ell$ range obtained varies between $\sim$1 mK$^2$ to $\leq$100 mK$^2$ for all three pointings across the entire range of angular modes. Despite being located quite far away from the galactic plane, these values may be sufficiently high to obscure the 21-cm signal (for which analytical calculations show a value $\sim$0.1 mK$^2$). The range of amplitude of the angular power spectrum for the residual data is slightly smaller than that obtained by \citet{arnab1}. They have done the analysis for $b = 44.89^\circ$, and the obtained values of $\mathcal{C}_\ell$ in the residuals vary between $\sim 10-100 mK^2$ for $1115 \leq \ell \leq 5083$ using direction-dependent approach. Previous works by \citet{bernardi09},  \citet{ghosh2011}, \citet{icobelli13}, \citet{tge17} were done for lower galactic latitudes (within $\pm$ 20$^\circ$), with the obtained values of $\beta$ varying between $\sim$1.8 to $\sim$2.9. This work demonstrates for the first time that the power-law index for diffuse emission is also in the same range of values for locations far away from the galactic plane. 

\section{Conclusion}
\label{conclusion}

Characterizing foregrounds is one of the most challenging steps in recovering the redshifted 21-cm signal targeted for studying CD and EoR. The foregrounds vary spatially across the entire sky, and hence more observation at low frequencies are essential for generating fiducial foreground models. This paper presents the results of such a low-frequency observation of the  Lockman Hole field located at very high galactic latitude. The data obtained from the GMRT archive is at 325 MHz covering almost 6 square degrees. A  mosaic image, $6^\circ \times 6^\circ$ across, has been produced after direction-dependent calibration of the data. The RMS level reached is $\sim$ 50 $\mu$Jy $\mathrm{beam}^{-1}$.  The sources recovered above $5\sigma_{\mathrm{RMS}}$ have been considered to produce a catalog of the field. The total sources recovered are 6186. Comparison has also been made with previous observation covering a part or whole of the same field to check for flux and position accuracy. The recovered fluxes and positions are found to be consistent with previous observations. Euclidean normalized source counts have been determined for the cataloged sources, after correcting for different errors and biases. The final counts are consistent with the previous observation for the same field as well as other parts of the sky.

The paper additionally probes variation in the power spectrum of DGSE at these frequencies for locations far off from the galactic plane. The angular power spectrum $\mathcal{C}_\ell$ as a function of $\ell$ has been determined for three different galactic locations of this field using TGE. The values of $\mathcal{C}_\ell$ lies between $\sim$1 mK$^2$ to $\sim$100 mK$^2$ for all the pointings. Despite being far-off from the galactic center, this is a very high value for $\mathcal{C}_\ell$ that can render recovery retrieval of the cosmological 21-cm difficult. Power law of the form $A\ell^{-\beta}$ has also been fitted in the angular power spectrum obtained. The fitted values for power-law index lie between 2.15 and 3.15 and also varies with varying the galactic latitude, thereby showing the necessity for more low-frequency observations for characterizing foregrounds. 


\section*{Acknowledgements}
The authors thank the staff of GMRT for making this observation possible. GMRT is run by National Centre for Radio Astrophysics of the Tata Institute of Fundamental Research. AM would like to thank Indian Institute of Technology Indore for supporting this research with Teaching Assistantship. AC would like to thank DST for INSPIRE fellowship. 
AM further acknowledges Ramij Raja, Sumanjit Chakraborty and Akriti Sinha for helpful discussions. 
This research made use of APLpy, an open-source plotting package for Python \citep{aplpy2012,aplpy2019} and Astropy,\footnote{\url{http://www.astropy.org}} a community-developed core Python package for Astronomy \citep{astropy:2013, astropy:2018}. 
The authors thank the anonymous reviewer and the scientific editor for helpful comments and suggestions that have helped to improve the quality of the work.



\bibliographystyle{mnras}
\bibliography{references} 




\bsp	
\label{lastpage}
\end{document}